# Information and Software Technology

# How to ask for technical help? Evidence-based guidelines for writing questions on Stack Overflow

Fabio Calefato[a], Filippo Lanubile[b], Nicole Novielli[b]

[a] *Dipartimento Jonico, University of Bari "A. Moro", via Duomo 259, 74123, Taranto, Italy*
[b] *Dipartimento di Infomatica, University of Bari "A. Moro", via E. Orabona 4, 70125, Bari, Italy*

**Keywords:**
Social software engineering
Crowdsourcing
Stack Overflow
Question answering
Sentiment analysis
Emotion mining

**ARTICLE INFO**

**ABSTRACT**

*Context:* The success of Stack Overflow and other community-based question-and-answer (Q&A) sites depends mainly on the will of their members to answer others' questions. In fact, when formulating requests on Q&A sites, we are not simply seeking for information. Instead, we are also asking for other people's help and feedback. Understanding the dynamics of the participation in Q&A communities is essential to improve the value of crowdsourced knowledge.

*Objective:* In this paper, we investigate how information seekers can increase the chance of eliciting a successful answer to their questions on Stack Overflow by focusing on the following actionable factors: affect, presentation quality, and time.

*Method:* We develop a conceptual framework of factors potentially influencing the success of questions in Stack Overflow. We quantitatively analyze a set of over 87 K questions from the official Stack Overflow dump to assess the impact of actionable factors on the success of technical requests. The information seeker reputation is included as a control factor. Furthermore, to understand the role played by affective states in the success of questions, we qualitatively analyze questions containing positive and negative emotions. Finally, a survey is conducted to understand how Stack Overflow users perceive the guideline suggestions for writing questions.

*Results:* We found that regardless of user reputation, successful questions are short, contain code snippets, and do not abuse with uppercase characters. As regards affect, successful questions adopt a neutral emotional style.

*Conclusion:* We provide evidence-based guidelines for writing effective questions on Stack Overflow that software engineers can follow to increase the chance of getting technical help. As for the role of affect, we empirically confirmed community guidelines that suggest avoiding rudeness in question writing.

## 1. Introduction

Software engineering involves a large amount of social interaction, as programmers often need to cooperate with others, whether directly or indirectly [13]. The massive adoption of social media has spurred the rise of the 'social programmer' [61] and the surrounding ecosystem [56]. In fact, social media have deeply influenced the design of software development-oriented tools such as GitHub[1] and Stack Overflow,[2] respectively, a social coding site and a community-based question answering site, which have recently attracted a considerable amount of research on social software engineering [15,60,62].

Stack Overflow is an example of an online community where social programmers do networking by reading and answering others' questions, thus participating in the creation and diffusion of crowdsourced knowledge and software documentation [1,62]. The success of Stack Overflow and, more in general, of community-based question-and-answer (Q&A) sites, mainly depends on the will of their members to answer others' questions. In fact, when formulating requests on Q&A sites, we are not simply seeking for information but we are also asking for other people's help and feedback. Understanding the dynamics of the participation in Q&A communities is essential to improve the value of crowdsourced knowledge [3,49,66]. Educating users to formulate questions properly is beneficial not only for the information seekers, since it increases the likelihood of receiving support, but also for the whole community, since it enhances effective knowledge-sharing behavior, also in the perspective of the creation of long-lasting value pieces of knowledge [3,11].

The Stack Overflow community provides official recommendations

---

*E-mail addresses:* fabio.calefato@uniba.it (F. Calefato), filippo.lanubile@uniba.it (F. Lanubile), nicole.novielli@uniba.it (N. Novielli).

[1] https://github.com
[2] http://stackoverflow.com

for effective question-writing that incorporates the suggestions provided by Skeet [57], the highest reputation community member, whose advices are considered the *de facto* standard by the community [36,59]. In this paper, we empirically validate a set of guidelines, based on both the community recommendations and the results from previous research in this field, against the findings provided by our empirical study, with the final goal of validating and incorporating such guidelines into an evidence-based netiquette. Specifically, we investigate how an information seeker can increase the chance of eliciting a *successful answer*,[3] that is, the answer marked as accepted by the question asker as the best one among those received.

To this aim, we develop a framework of factors influencing the success of questions in Stack Overflow. Specifically, we focus on actionable factors that can be acted upon by software developers when writing a question to ask for technical help, namely *affect* (i.e. the positive or negative sentiment conveyed by text), *presentation quality*, and *time*. The asker *reputation* is also included as a control factor.

To fulfill our research goal, we follow a mixed-methods approach characterized by a sequential explanatory strategy [19]. We analyzed a dataset of 87 K questions extracted from the official Stack Overflow dump. We combined: (1) a logistic regression analysis, to estimate the probability of success of a question based on affect and the other actionable factors in our framework, as discussed next; (2) a qualitative analysis of the questions conveying either positive or negative sentiment, to complement the findings of the previous step; and (3) a user survey, to garner additional qualitative insight into the users' perception of the question-writing guidelines.

The main contributions of this paper are defined as follows. First, we provide the definition of a general framework of the technical, linguistic, and human factors that predict the probability of receiving a successful answer in Stack Overflow. Second, we provide an empirical assessment of how sentiment polarity correlates to the success of Stack Overflow questions. Our findings about the role of affect in Stack Overflow questions represent a novel contribution of this study. Finally, starting from prior research findings and the *de facto* standard recommendations adopted by the Stack Overflow community on how to write good questions, we identify those that are supported by empirical validation and, thus, we provide a set of evidence-based guidelines for effective question writing.

The remainder of this paper is organized as follows. In Section 2, we describe the conceptual framework that informs the empirical study, as well as the metrics that operationalize each factor of success. Section 3 describes the Stack Overflow dataset whereas Section 4 reports the results of our empirical investigation. In Section 5, we provide the empirical guidelines derived from our study. Finally, in Section 6 we discuss limitations and in Section 7 we provide conclusions, discuss open challenges, and propose future research work.

## 2. Conceptual framework

In the following, we develop a conceptual framework for the analysis of factors influencing the success of questions in Stack Overflow, that is, the probability for a question to receive one answer that is accepted as a solution by the asker. The framework is built upon the evidence provided by previous studies on successful questions in Q&A sites, which highlighted those factors that can influence, to a different extent, the chances of receiving help from other community members. Furthermore, we consider community recommendations as provided by Stack Overflow.

### 2.1. Background and related work

Jon Skeet is the Stack Overflow contributor with the highest reputation since 2009. He has provided suggestions about how to write good questions that can attract more contributions from potential helpers [57]. Skeet's recommendations are considered the *de facto* standard by the Stack Overflow community. As such, they are integrated into the official guidelines of the website help center [58]. Furthermore, they are often cited in Meta, a Q&A site where users discuss features and issues regarding Stack Exchange, the network of Q&A sites originated by the success of Stack Overflow [35]. Such guidelines are complemented by evidence provided by recent empirical studies about the influence of technical, linguistic, and human aspects of question-writing on the success of questions in Q&A sites.

Technical aspects, which mainly depend on presentation quality, are known to be among the driving factors of the success of requests in Q&A sites [4,64]. In fact, Asaduzzaman et al. [4] showed that unclear and vague questions remain unanswered in Stack Overflow. Instead, the presence of example code snippets in questions is positively associated with their success [4,17], especially in the case of code reviews [57,64]. Besides, research has recently begun to investigate linguistic aspects too, by looking at how lexicon [37] and narratives [2] in help requests may influence their success. The investigation on human factors has been restrained to analyzing the effect of people's expertise and degree of involvement in the community, suggesting as social reputation [2,3] and personality [6] play a role in the success of online requests.

A recent trend has emerged to study the role of affect in software engineering [55], thus highlighting the importance of sentiments in software development [24,26,38,39,41], maintenance, and evolution [25,31]. However, we know very little about the influence of expressing emotions when asking for technical help in Stack Overflow although there are clues that it can be important. For instance, new users often complain about harsh posts from expert contributors [35]. Likewise, Stack overflow [58] and Skeet's guidelines [57] invite both information seekers and providers to be patient and polite, avoiding rudeness, especially while interacting with new users. Furthermore, empirical evidence provided by previous research in related domains advocates in favor of the consideration of affect as a dimension of our analysis. Calefato et al. [8] analyzed the use of affective lexicon in four non-technical Q&A sites of the Stack Exchange network (Mathematics, Bitcoin, Arqade and Science Fiction). Their study reveals that the effect of sentiment and gratitude expressions on the success of a question may vary depending on the community being analyzed. Kucuktunc et al. [28] performed a large-scale sentiment analysis study on Yahoo! Answers. Albeit not focused on questions, their work shows that best answers tend to a neutral sentiment, thus suggesting that expressing either positive or negative sentiment in Q&A sites might reduce the perceived quality of a post. Althoff et al. [2] found that expressing gratitude upfront in Reddit is positively correlated with the success of requests because is seen as a clue of positive disposition towards a potential answerer. As such, we include emotion lexicon in questions as an additional source of information in our analysis. Here, in particular, we focus on the influence that emotion polarity – the positive or negative emotional style of questions – has on the chance of eliciting a successful answer on Stack Overflow.

### 2.2. Critical factors of successful questions

Our conceptual framework illustrated in Fig. 1, depicts three critical and actionable success factors reported in the literature so far – namely *affect, presentation quality*, and *time* – which can be acted upon by an asker when writing a question. The asker's *reputation* is also added as a control dimension, due to the evidence of its impact on the success of requests in online communities.

**Affect.** Affective states range from long-standing features, such as personality traits, to more transient ones, such as emotions [54].

---

[3] Hereinafter, *successful answer* and *best answer* are used interchangeably to refer to the answer accepted as the solution by the asker; hence, a *successful question* is one that has received a successful answer.

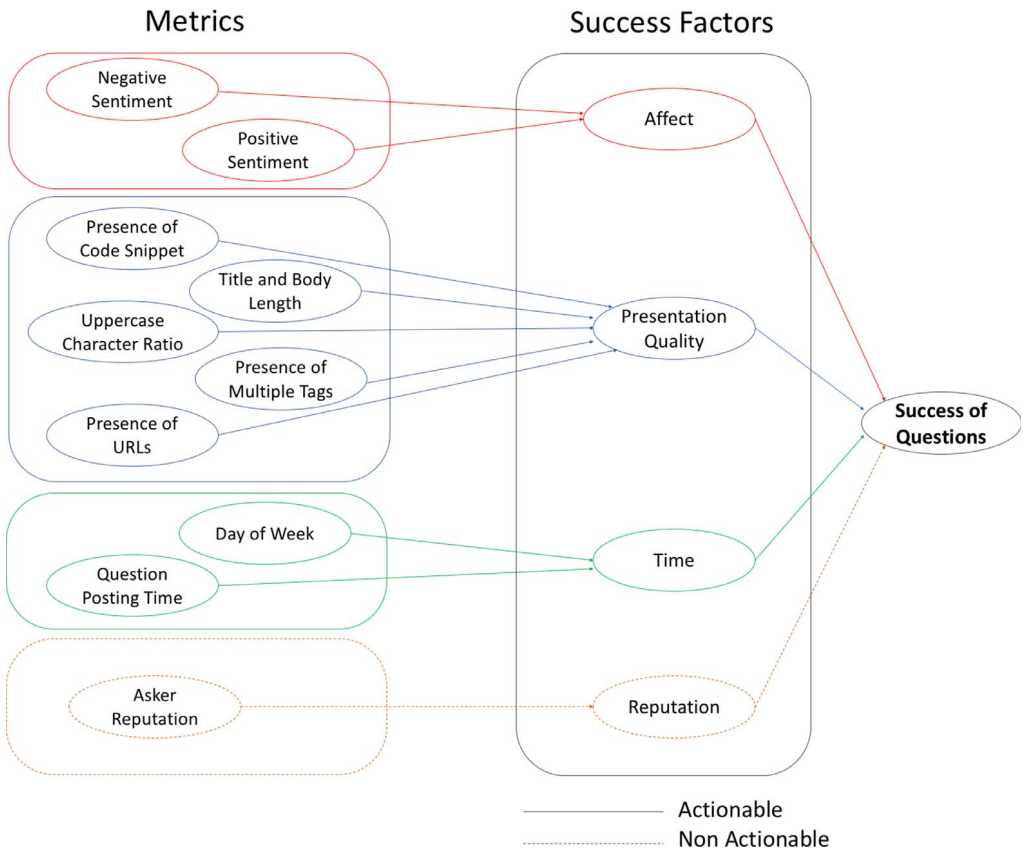

Fig. 1. Conceptual framework of critical success factors of questions in Stack Overflow.

Mining affective states from text [12] involves, on one hand, to model them according to bi-dimensional models representing the affect polarity (i.e., the overall positive, negative or neutral orientation) and its arousal (i.e., the level of activation) [53]; on the other hand, some studies explicitly deal with discrete emotion labeling of text, by looking for linguistic cues of specific affective states (e.g., Ekman's six basic emotions: anger, disgust, fear, happiness, sadness, and surprise) [18].

Existing evidence about requests for technical help suggests that expressing sentiment, either positive or negative, might decrease the chances of getting help. More specifically, in studying the personality of Stack Overflow users, Bazelli et al. [6] discovered that the authors of downvoted questions exhibit lexical cues of extroversion. Furthermore, the Stack Overflow guidelines explicitly suggest avoiding rudeness, harassment, and bullying [58]. Consistently, Skeet suggests avoiding also expressions carrying positive sentiment polarity, such as greetings and sign-offs, because these can be perceived as a distraction and, therefore, they tend to be edited out by other users with moderation privileges [57].

While we observe contrasting evidence about the impact on sentiment in non-strictly technical domains [2,8,28], guidelines provided by the Stack Overflow community advocate in favor of a neutral style and even discourage the expression of positive sentiment. Therefore, through our investigation, we expect to find that expressing sentiments in a question in Stack Overflow does have an influence on its probability of eliciting a successful answer. Whether this influence is positive or negative is what we seek to determine.

**Presentation quality.** Previous research has found that providing information in a manner that promotes readability and comprehension is regarded as the strongest indicator of the quality of a question [4,17,64,57]. As further confirmation of the importance of presentation quality, Stack Overflow uses a set of simple textual metrics (e.g., length, use of uppercase, presence of URLs) to feed a queue of supposed low-quality questions, which are then analyzed by a selected subset of high-reputation users with moderation privileges. These metrics have also been previously used by Ponzanelli et al. [49] to develop an automated approach for the identification of high- vs. low-quality questions in Stack Overflow. While the other text-based metrics are the object of SO community guidelines (e.g., be concise, use a short title) [57], Ponzanelli et al. do not clarify whether the presence of URLs is an indicator of a poor- rather than a high-quality question. Yet, we argue that, as in the case of defining multiple tags, adding URLs to external resources may help give further context to a help request. Finally, Skeet suggests using capital letters only where appropriate, as uppercase might have a negative impact on the readability of text [57].

Given the considerable body of evidence, through our analysis, we expect to determine which metrics are actually capable of evaluating the presentation quality of questions in Stack Overflow.

**Time.** Bosu et al. [7] have found that the most successful time slices correspond to the working time in the USA, where most of Stack Overflow experts resided at the time of their study. They observed that the highest proportion of questions with accepted answers to be in the time frame corresponding to 3:00–6:00 PM of West Coast US time, that is, when most experts were available (∼40% of all Stack Overflow users, at the time of their study). In addition, they found that questions posted during the weekend are more likely to be answered than questions posted during the week.

Through our analysis, we seek to confirm the existing evidence that there are time windows when questions are more likely to be successfully answered.

**Reputation.** The users' status, expressed through reputation score, resulted among the best predictors of success of requests in online communities [2]. In their research, the users with higher status in the community object of the study (Reddit.com) were more likely to receive the help requested. Conversely, in the study by Treude et al. [64], questions posted by novices seems to be more frequently answered than others in Stack Overflow, probably because they are easier to answer. In our previous research [9], we observed a positive correlation between the asker's reputation score and the probability of an answer being accepted. Seeking an explanation, we studied the distribution of the badge 'Scholar' among all users who have asked at least a question. Users unlock this badge when accepting an answer for the first time. We found that only the 29% of novice users holds the scholar badge. The percentage increases with the reputation score (up 99% for top reputation users) thus indicating a tendency of less expert users to not mark successful answers as accepted. Either way, user reputation has been demonstrated to be correlated with the success of a post. Therefore, although not actionable, we include asker *reputation* as a control variable.

**Table 1**
Examples of sentiment expression in Stack Overflow questions with associated scores as issued by Senti Strength and discretized as Boolean.

| | Sentiment Strength Scores | | Discretized Sentiment Scores | | |
|---|---|---|---|---|---|
| Excerpts from the Stack Overflow dataset | Positive | Negative | Positive | Negative | Neutral |
| "I have very simple and stupid trouble […]. I'm pretty confused, explain please, what is wrong?" | +1 | −2 | False | True | False |
| "[…] Any help would be really great! :-)" | +5 | −1 | True | False | False |
| "I want them to resize based on the length of the data they're showing" | +1 | −1 | False | False | True |

*2.3. Measuring the factors*

In the following, we operationalize the factors of the conceptual framework by reusing, whenever possible, the relevant metrics already defined in previous studies.

**Affect**. Consistently with previous research on affect in software engineering [16,22,24,25,44,47,48,50], we rely on sentiment analysis [46] to detect the polarity and strength of the sentiment conveyed in the body of a question. As such, we operationalize the *affect* factor through the following two metrics.

*1. Positive Sentiment and 2. Negative Sentiment*

This metric represents the overall sentiment orientation of a question. Specifically, we measure the overall positive (*Positive Sentiment*) and negative (*Negative Sentiment*) polarity of the sentiment expressed in a question body. To capture the sentiment of questions, we used SentiStrength [63], a sentiment analysis tool already employed in social computing [23,28,63], which is capable of dealing with short informal text, including abbreviations, intensifiers, and emoticons. Based on the assumption that a sentence can convey mixed sentiment, SentiStrength outputs both positive and negative sentiment scores for an input text. Positive sentiment scores range from +1 (absence of positive sentiment) to +5 (extremely positive) while negative sentiment scores range from −1 (absence of negative sentiment) to −5 (extremely negative). In Table 1, we report a few examples of sentiment expression from our dataset with associated sentiment scores as issued by SentiStrength. Consistently with previous research [2], we discretize sentiment scores. In particular, we treat *Positive Sentiment* and *Negative Sentiment* as Boolean variables by encoding whether a question shows in its lexicon a 'positive' (positive score in [+2, +5]) or 'negative' (negative score in [−2, −5]) affective load. A question is 'neutral' when positive sentiment score = +1 and negative sentiment score = −1.

**Presentation Quality.** We measure the presentation quality of a question by using the following metrics.

*1. Title length and 2. Body length*

Both metrics are calculated as number of words. Too short questions may remain unanswered, failing to clarify their meaning to community members [4]. On the other hand, overly long questions may be associated with a greater effort in providing an answer and might discourage potential helpers. Consistently with previous research [2], we choose to discretize both *Title Length* and *Body Length* to allow meaningful assessment of the effect size of such variables in our analysis. In particular, we apply clustering-based discretization of both metrics using k-means clustering as implemented in the *arules* package for R [27]. We identify three clusters for *Body Length*, namely *Short* (length in characters in [0, 90[), *Medium* ([90, 200[characters), and *Long* (longer than 200 characters), as well as for Title Length, namely *Short* ([0, 6[characters), *Medium* ([6, 10[characters), and *Long* (longer than 10 characters).

*3. Presence of URLs*

We check for the presence of links in a question and store the results in a binary, yes/no variable. We include the *Presence of URLs* in the set of presentation quality metrics because links are typically embedded to provide a richer context, i.e., providing evidence of previous searches on the site or adding extra information useful for a full comprehension of the question.

*4. Presence of multiple tags*

Stack Overflow allows users to search and browse questions based on user-specified tags. When properly employed, tags associated with a question are useful to attract potential answerers based on the topic of interest. Specifying at least one tag is mandatory when asking a question on Stack Overflow. Then, we use a binary metric to model the presence of single vs. multiple tags.

*5. Uppercase character ratio*

The abuse of uppercase is often an indicator of low-quality posts, often contributed by community newcomers. Furthermore, it severely impacts readability [57]. This metric is calculated as the ratio between the total uppercase characters over the total characters in a question. As in the case of *Title* and *Body Length*, we discretize the *Uppercase Character Ratio* using the *arules* package, resulting in two clusters, namely *Low* uppercase ratio (values in [0, 0.10[) and *High* uppercase ratio (more than 0.10).

*6. Presence of code snippet*

A binary, yes/no metric, indicating whether a question embeds an excerpt of code within the HTML ⟨code⟩ block, provided as an example by the asker. Code snippets are considered key to success in 'code review' questions [64].

**Time.** We operationalize the time factor in terms of the following couple of metrics:

1. Question posting time (GMT), coded with values in {Morning, Afternoon, Evening, Night}.
2. Day of Week, coded with values in {Weekend, Weekday}

**Reputation.** We operationalize the reputation factor in terms of the:

*1. Asker reputation*, that is the reputation of the author of the question at the time it was posted on Stack Overflow, as retrieved from the dump. In our model we represent the reputation as a categorical variable based on how Stack Overflow reputation system assigns users to the following categories, given the reputation score gained: *New* users (score < 10), *Low Reputation* users (score in [10, 1000[), *Established* users (score in [1000, 20K[), and *Trusted* users (score ≥ 20 K).

**3. Dataset**

To date, Stack Overflow counts more than 6 M users providing 21 M answers to 13 M questions about computer programming.[4] The dataset employed in this study is obtained from a dump of all user-contributed

---
[4] Last accessed on January 2017

Table 2
The dataset of Stack Overflow questions employed for the current study, broken down by factors and variables.

| Questions | | | N | Asker reputation | | | |
|---|---|---|---|---|---|---|---|
| | | | | New | Low reputation | Established | Trusted |
| **Overall** | | | 87,373 | 23,433 | 53,637 | 9927 | 376 |
| **Resolved** | | | 30,797 (35%) | 4853 (21%) | 21,400 (40%) | 4362 (44%) | 182 (48%) |
| **Affect** | with positive sentiment | | 22,379 (26%) | 6925 (29%) | 13,506 (25%) | 1888 (19%) | 60 (16%) |
| | with negative sentiment | | 9578 (11%) | 2577 (11%) | 5758 (11%) | 1186 (12%) | 57 (15%) |
| **Time** | Day of week | Weekend | 14,035 | 3939 | 8475 | 1561 | 60 |
| | | Weekday | 73,338 | 19,494 | 45,162 | 8366 | 316 |
| | Question posting time (GMT) | Morning | 25,863 | 7073 | 15,906 | 2798 | 86 |
| | | Afternoon | 29,074 | 7655 | 17,933 | 3358 | 128 |
| | | Evening | 17,947 | 4728 | 10,961 | 2176 | 82 |
| | | Night | 14,489 | 3977 | 8837 | 1595 | 80 |
| **Presentation Quality** | Presence of code snippet | True | 60,591 | 13,927 | 38,509 | 7845 | 310 |
| | | False | 26,782 | 9506 | 15,128 | 2082 | 66 |
| | Uppercase character ratio | Low | 82,285 | 21,765 | 50,687 | 9472 | 361 |
| | | High | 5088 | 1668 | 2950 | 455 | 15 |
| | Body length | Short | 60,112 | 16,197 | 16,197 | 6912 | 227 |
| | | Medium | 23,029 | 6027 | 6027 | 2557 | 113 |
| | | Long | 4232 | 1209 | 1209 | 458 | 36 |
| | Title length | Short | 7368 | 2099 | 2099 | 673 | 35 |
| | | Medium | 40,582 | 10,568 | 10,568 | 4513 | 146 |
| | | Long | 39,423 | 10,766 | 10,766 | 4741 | 195 |
| | Presence of multiple tags | True | 75,975 | 19,733 | 19,733 | 8629 | 332 |
| | | False | 11,398 | 3700 | 3700 | 1298 | 44 |
| | Presence of URLs | True | 15,608 | 3819 | 3819 | 2290 | 111 |
| | | False | 71,765 | 19,614 | 19,614 | 7637 | 265 |

Stack Overflow content from July 2008 to September 2014.

To foster community engagement, the social reputation system of Stack Overflow is designed to incentivize contributions and allow assessment of the trustworthiness of users. Each release of the Stack Overflow dump reports the reputation score of each user at the moment the dump is created. Since the reputation score of users in the Stack Overflow community evolves over time, we cannot use this metric in our data dump as a measure of the status of a user. On the contrary, we need to assess the reputation of the users at the time they pose a given question in our dataset. Since Stack Overflow allows users to gain at most 200 reputation points per day, it is reasonable to assume that the reputation category of most users stays unvaried over a month. Therefore, we built the dataset for our analysis by considering the questions posted during the last month of the dump (14th August − 14th September 2014). From this set, we removed self-answered questions, which would bias the results. Furthermore, we excluded questions that were removed or closed by moderators. Analogously to Ponzanelli et al. [49], we removed the questions that were edited after the original post. Due to a limitation in the information provided by the dump, we cannot know how a question edited afterward looked like when the accepted answer was provided. As such, this step is necessary since modifying a question may have indirect side effects on the quality evaluation. Finally, we removed questions that were posted up to three days before the creation date of the dump. This is a necessary step to avoid introducing a bias in the results because the community did not have the time to answer them by the time the dump was created. We specifically chose three days as a cutoff point because we observed that the resolved questions in the dump follow a long-tailed distribution, with those resolved within the third day accounting for over 92%. Hence, the final dataset resulting from this pre-processing phase ended up to containing 87,373 questions.[5]

Questions can have at most one accepted answer, chosen by the asker if she thinks that it solves the problem. In line with previous research [64], we consider as 'successful' those questions for which an accepted answer has been provided. As a result, the final dataset contains 30,797 successful questions (35%). Table 2 reports a breakdown of the questions, thus providing a characterization of the dataset across each variable of our framework.

Before calculating all the text-based metrics (*Sentiment* scores, *Title* and *Body Length*, *Uppercase Character Ratio*), we pre-processed questions to discard code snippets and remove HTML tags.

Since an excessive correlation among the predictors may hide their individual contributions to the model, we performed a collinearity analysis of the explanatory variables in the dataset. The Variance Inflation Factor (VIF) [52] is a measure of how much the variance of an estimated regression coefficient increases if the predictors are correlated. The higher the VIF index for a predictor, the greater its degree of collinearity with the other explanatory variables. As a rule of thumb, a VIF > 10 indicates strong evidence of collinearity [52,33]. However, some researchers (e.g., [14]) adopt a stricter criterion, referring to values higher than five to denote a high collinearity. In Table 3, we report the VIF index for all the predictors in the model. Because VIF values for all predictors in our model are close to 1, we can confirm that the variance of each explanatory variable is not affected by the value assumed by the others.

---

[5] The dataset is available for research purposes and can be downloaded from: https://goo.gl/whZEWA

Table 3
Assessing the degree of collinearity for the framework predictors (VIF values in [5, 10] denote high collinearity).

| Factor | Predictor | VIF |
|---|---|---|
| Affect | Positive sentiment | 1.06 |
| | Negative sentiment | 1.04 |
| Time | Day of week | 1.01 |
| | Question posting time (GMT) | 1.36 |
| Presentation quality | Presence of code snippet | 1.01 |
| | Uppercase character ratio | 1.01 |
| | Body length | 1.11 |
| | Title length | 1.01 |
| | Presence of multiple tags | 1.02 |
| | Presence of URLs | 1.01 |
| Reputation | Asker Reputation | 1.20 |

Table 4
Results of logistic regression on the complete dataset.

| | Factor | Predictor | Coefficient estimate | Odds Ratio | p-value | Sig. |
|---|---|---|---|---|---|---|
| *Actionable Factors* | Affect | *Positive Sentiment in the Question Body* ('Neutral' as default value) | −0.06 | 0.94 | .0003 | *** |
| | | *Negative Sentiment in the Question Body* ('Neutral' as default) | −0.21 | 0.81 | <2e−16 | *** |
| | Time | *Day of Week* ('Weekday' as default) | | | | |
| | | Weekend | 0.10 | 1.10 | 4.50e–07 | *** |
| | | *GMTHour* ('Morning' as default) | | | | |
| | | Afternoon | 0.07 | 1.07 | .0004 | *** |
| | | Evening | 0.15 | 1.16 | 3.68e−12 | *** |
| | | Night | 0.13 | 1.14 | 3.53e−09 | *** |
| | Presentation Quality | *Presence of Code Snippet* ('No' as default) | 0.71 | 2.04 | <2e−16 | *** |
| | | *Uppercase Character Ratio* ('High' as default) | | | | |
| | | Low Uppercase Ratio [0, 0.1[ | 0.12 | 1.27 | <2e−16 | *** |
| | | *Body Length* ('Short' as default, where 'Short' is in [0, 90[) | | | | |
| | | Medium [90, 200[ | -0.21 | 0.81 | <2e−16 | *** |
| | | Long [200, +∞[ | -0.42 | 0.65 | <2e−16 | *** |
| | | *Title Length* ('Short' as default, where 'Short' is in [0, 6[) | | | | |
| | | Medium [6, 10[ | 0.05 | 1.05 | .0595 | |
| | | Long [10, +∞[ | 0.05 | 1.05 | .0465 | * |
| | | *Presence of Multiple Tags* ('Single tag' as default) | -0.12 | 0.88 | 1.10e-08 | *** |
| | | *Presence of URLs* ('No' as default) | -0.01 | 0.99 | .4529 | |
| *Control Factor* | Reputation | *Asker Reputation* ('New' as default value, where 'New' is in [0,10 [) | | | | |
| | | Low [10, 1K[ | 0.87 | 2.39 | <2e−16 | *** |
| | | Established [1 K, 20K[ | 0.98 | 2.68 | <2e−16 | *** |
| | | Trusted [20 K, +∞[ | 1.17 | 3.23 | <2e−16 | *** |
| | | (Intercept) | -1.60 | – | <2e−16 | *** |
| | | *Significance codes: '***' 0.001 '*' 0.05* | | | | |

## 4. Results

We conduct our analysis by applying a logistic regression for estimating the extent to which the actionable factors *affect, time*, and *presentation quality* correlate to the probability of success of a question. Even if not actionable, we include the asker *reputation* as a control factor. We treat the success of a question as the dependent variable and the metrics associated with the four factors as the independent variables. We use logistic regression for its ease of interpretability since it allows us to reason about the significance of one factor given all the others. We perform the analysis using the R statistical software environment. Results of logistic regression analysis are reported in Section 4.1. To understand the role played by affective states in the success of Stack Overflow questions, we complement the correlation analysis with the qualitative analysis of positive and negative emotions in our dataset, as reported in Section 4.2. Finally, in Section 4.3 we present the results of a survey that we conducted to understand the community members' perception of suggestions incorporated in the final guidelines.

### 4.1. Which factors are predictive of success?

The results of the logistic regression are reported in Table 4, with predictors grouped by factor (control vs. actionable). For each predictor, we report the coefficient estimate, the odds ratio, the p-value, and the statistical significance. A logistic model is said to provide a better fit to the data if it improves over the *null* model (i.e., the intercept-only baseline model, where no predictors are included). The *null* model predicts all the questions as 'unsuccessful', which is the majority class in our dataset. As suggested by Menard [33], we measure the improvement over this baseline using the likelihood ratio test. Results are reported in the last row of Table 4 and show that our logistic model is significantly different from the *null* model ($p < .05$). Furthermore, we assess the goodness of fit of our model by calculating the AUC value of the ROC curve (AUC = 0.65).[6]

In a logistic model, the sign of the coefficient estimate indicates the positive/negative association of the predictor with the success of a question. The odds ratio (OR) weighs the magnitude of this impact: the closer its value to 1, the smaller the impact of the parameter on the chance of success. A value lower than 1 corresponds to a negative association of the predictor with success (negative sign of the coefficient estimate), and the opposite for a value higher than 1. An OR = $x$ technically implies that the odds of the positive outcome are $x$ times greater than the odds of the negative outcome. We would like to remind the reader that odds ratio is an asymmetrical metric: positive odds can vary from 1.0 to infinity while decreasing odds ratios are bounded by 0 [45]. In this sense, we cannot directly compare the magnitude of effect size for predictors with increasing ORs with those with decreasing ORs, as in the case of *Asker Reputation = Trusted* (OR = 3.23) and *Long Body* (OR = 0.65).

We observe that the *Asker Reputation* shows the most significant effect size in our model. In our previous research, we observed that new users may not be familiar with the community rules involving the acceptance of the best answer [9]. To further our understanding of these results, two researchers analyzed the question threads originated by 100 randomly selected unsuccessful questions posted by *New* users, from which they extracted the following coding schema of the main reasons why the acceptance vote is missing:

- *No answers*, indicating questions for which no answers have been posted;
- *No useful answers*, indicating questions that received answers but none of them was judged as potentially useful to the asker;
- *Useful answers*, indicating questions that received either one or more *potentially useful answers* in the thread, based on the review of their content and the number of upvotes received, or a *solving answer*, according to the asker's comments that explicitly identify the best answer that was supposed to be accepted.

The annotation was performed by manually inspecting each extracted question thread in the dataset to identify the accepted solution (if any) among the existing answers. Two researchers first performed the annotation independently; then, they compared the results and iteratively resolved conflicts until complete agreement was reached.

---

[6] We experiment with other classification algorithms and observed comparable performance.

Fig. 2. An example of an unsuccessful question from our dataset.

> SSRS Report Grouping and Sums
>
> **Am struggling** getting to terms with SSRS and have a report to write that I cannot seem figure out how to do. Basically I have I hierarchy of Parent Groups that are related Bldg > Floor > Cost Code and fields outside of these groups for Sum(area), sum(seats) that calculate the area and seats on each floor belonging to each cost code...simple so far. The problem I have is I need to calculate the percentage of area and seats each cost code takes up from the total on a floor and in the building.
>
> By putting calculations in the group section for building total and floor total I can refer to the value in the text box and perform calculations from it.....great, but I cannot hide the columns as they are within the group section! Also going to the column properties I can hide the column but then get white space in the report where the column was**......not very useful!**
>
> How can I sum a total area and seats for a floor and a building so I can use it to calculate a percentage for each cost code while still showing the totals spilt by cost code?

The results are reported in Table 5 and show the tendency of new users to not accept answers. In fact, we found that 38% of questions do not have an accepted answer albeit a solving solution was provided.

As for *affect*, we observe a negative correlation with success for both *Positive* (OR = 0.94) and *Negative Sentiment* (OR = 0.81). As for *presentation quality*, the appropriate use of uppercase in the question body is positively correlated with success (OR = 1.27). The chance of getting an accepted answer decreases as the length of the question body increases (OR = 0.65 for questions longer than 199 words). On the contrary, the use of code snippet is positively associated with success (OR = 2.04). Finally, in relation to *Time*, we observe an increased chance of success in GMT evening hours (OR = 1.16).

The logistic regression parameters correspond to changes in log-odds space rather than probability space and, therefore, they may not be straightforward to interpret. Therefore, to further our understanding of how to ask a question on Stack Overflow, in the following we analyze one unsuccessful question selected from our dataset. In particular, we show a real example of how a novice user can increase the probability of getting help by controlling the actionable factors in our framework.

The unsuccessful question in Fig. 2 is a long question (> 200 words) posted by a newbie (i.e., belonging to the *New* user category), for which expressions of negative sentiment are found (shown in bold in the Fig. 2), but no code snippet is provided. Based on the logistic regression model built from our data, the coefficients of which are shown in Table 4, we estimate for the question an initial probability of success of $p = .09$ (see the first row for column *New* users in Table 6), which is consistent with its actual, unsuccessful status. The other rows in Table 6 show how to improve the initial chance of success of the original question by controlling the actionable factors in our framework. More in details, by observing a neutral sentiment score (i.e., setting both *Positive* and *Negative Sentiment* predictors to 'neutral') the probability of success would increase to $p = .11$. By adopting a more concise writing style (i.e., setting the predictor *Body Length* value to 'short'), the chance of success would further increase to $p = .16$. By also adding a code snippet to better clarify the problem (i.e., setting the *Presence of Code Snippet* value to 'yes'), the asker would further increase the chance of success to $p = .28$.

We repeat the simulation also for the other user reputation categories, namely *Low Reputation, Established*, and *Trusted* users. The higher probability estimates observed in such cases (see values in the related columns in Table 6) indicate how the asker's reputation remains the predictor with the stronger correlation with the success of questions.

We observe a prevalent effect size of asker *reputation* (i.e., the control factor) from the coefficient estimates in our model (see Table 4) and the previous simulations (see Table 6). Therefore, it becomes important to assess the impact of the actionable factors for each reputation category of users. Thus, we repeated the logistic regression analysis by splitting the original dataset into four subsets, based on the reputation categories of the asker. Results are reported in Table 7 grouped by factor, to enable comparison between reputation groups. Again, the *p*-values and statistical significance (below the *p*-value) for each predictor are reported.

We observe how *Presentation Quality* is the most relevant success factor. In particular, the *Presence of Code Snippet* is strongly and positively correlated with success for all reputation categories. On the contrary, the *Uppercase Character Ratio* is negatively associated with success for all reputation categories, except for *Trusted* users, for whom statistical significance is not observed. As for *Body Length*, the higher the reputation the stronger the negative correlation with success. Conversely, *Title Length* appears to have a statistically significant correlation with success only for *New* reputation users, albeit the association is not strong (OR equal to 1.14 and 1.13 for 'Medium' and 'Long' titles, respectively). As far as *Affect* is concerned, the results confirm the evidence provided by the first analysis (see Table 4), indicating that *Negative Sentiment* is negatively associated with success in Stack Overflow, with effect size particularly strong for *Trusted* users (OR = 0.55).

Table 5
An analysis of 100 unsuccessful answers posted by new users.

| Actual question status | % |
| --- | --- |
| No answers | 44% |
| No useful answers | 18% |
| Useful answers, of which: | 38% |
| Potentially useful answers | 21% |
| Solving answer | 17% |

Table 6
An example of how a user can control the actionable factors to increase the probability of success for a question.

| | Estimated probability of success by user category | | | |
| --- | --- | --- | --- | --- |
| Question status | New | Low reputation | Established | Trusted |
| Initial probability | $p = .09$ | $p = .18$ | $p = .20$ | $p = .23$ |
| Adopting a neutral style | $p = .11$ | $p = .23$ | $p = .25$ | $p = .28$ |
| Being concise (short question) | $p = .16$ | $p = .31$ | $p = .33$ | $p = .38$ |
| Adding a code snippet | $p = .28$ | $p = .48$ | $p = .51$ | $p = .55$ |

Table 7
Results of logistic regression by reputation category.

| Reputation | New | | | Low | | | Established | | | Trusted | | |
|---|---|---|---|---|---|---|---|---|---|---|---|---|
| Predictor | Coeff. Estim. | Odds Ratio | p-value | Coeff. Estim. | Odds Ratio | p-value | Coeff. Estim. | Odds Ratio | p-value | Coeff. Estim. | Odds Ratio | p-value |
| *Affect* | | | | | | | | | | | | |
| Positive Sentiment ('Neutral' as default) | 0.00 | 1.00 | .9371 | −0.08 | 0.93 | .0003 *** | −0.13 | 0.88 | .0154 * | 0.29 | 1.34 | .3505 |
| Negative Sentiment ('Neutral' as default) | −0.23 | 0.79 | 5.72e − 05*** | −0.21 | 0.81 | 1.63e − 12 *** | −0.13 | 0.88 | .0461 * | −0.68 | 0.51 | .0364 * |
| *Time* | | | | | | | | | | | | |
| Weekend ('Weekday' as default) | 0.10 | 1.11 | .0179 * | 0.12 | 1.13 | 8.98e − 07 *** | 0.00 | 1.00 | .9747 | −0.20 | 0.82 | .5005 |
| *GMTHour ('Morning' as default)* | | | | | | | | | | | | |
| Afternoon | 0.03 | 1.03 | .4360 | 0.08 | 1.08 | 8.98e − 07 *** | 0.05 | 1.05 | .3990 | 0.14 | 1.15 | .6137 |
| Evening | 0.12 | 1.13 | .0106 * | 0.18 | 1.19 | .0009 *** | 0.02 | 1.02 | .7377 | 0.42 | 1.53 | .1919 |
| Night | 0.15 | 1.17 | .00214 ** | 0.13 | 1.14 | 2.09e − 11 *** | 0.07 | 1.07 | .3189 | 0.49 | 1.64 | .1302 |
| *Presentation Quality* | | | | | | | | | | | | |
| Presence of Code Snippet ('False' as default) | 0.78 | 2.18 | < 2e − 16 *** | 0.70 | 2.02 | < 2e − 16 *** | 0.63 | 1.88 | < 2e − 16 *** | 0.60 | 1.83 | .0378 * |
| *Uppercase Char. Ratio ('High' as default)* | | | | | | | | | | | | |
| Low [0,0.10[ | 0.21 | 1.23 | 5.70e − 08 *** | 0.19 | 1.21 | < 2e − 16 *** | 0.16 | 1.73 | .0014 ** | −0.12 | 0.89 | .6449 |
| *Body Length (Short as default)* | | | | | | | | | | | | |
| Medium [90, 200[ | −0.02 | 0.98 | .5992 | −0.22 | 0.80 | < 2e − 16 *** | −0.43. | 0.65 | < 2e − 16 *** | −0.58 | 0.56 | .0236 * |
| Long [200, + ∞[ | −0.28 | 0.76 | .0007 *** | −0.45 | 0.63 | < 2e − 16 *** | −0.45 | 0.64 | 1.50e − 05 *** | −1.16 | 0.31 | .0058 ** |
| *Title Length (Short as default)* | | | | | | | | | | | | |
| Medium [6,10[ | 0.13 | 1.14 | .0294 * | 0.04 | 1.04 | .2615 | −0.03 | 0.97 | .7103 | 0.23 | 1.25 | .5709 |
| Long [10, + ∞[ | 0.12 | 1.13 | .0351 * | 0.04 | 1.04 | .2120 | −0.01 | 0.99 | .9380 | −0.15 | 0.86 | .6859 |
| Presence of Multiple Tags ('Single tag' as default) | 0.14 | 1.15 | .0032 ** | −0.18 | 0.84 | 1.16e − 10 *** | −0.36 | 0.70 | 2.92e − 09 *** | 0.21 | 1.24 | .5202 |
| Presence of URLs ('False' as default) | −0.10 | 0.91 | .0319 * | −0.02 | 0.98 | .4830 | 0.09 | 1.10 | .0614 | 0.32 | 1.37 | .1782 |
| (Intercept) | −2.26 | — | < 2e − 16 *** | −0.93 | — | < 2e − 16 *** | −0.43 | — | .0002 *** | −0.78 | — | .2539 |

*Significance codes:* '***' 0.001 '**' 0.01 '*' 0.05

**Table 8**
Coding schema for emotions and their direction.

| | |
|---|---|
| Sentiment Direction | **Object:** The sentiment is expressed towards an object, such as a technical issue<br>e.g.:<br>• "I'm aware this code is really awful"<br>• "This is irritating for users"<br>Mapping with categories of the OCC model [43]:<br>• Aspect of objects {liking/disliking, love/hate}<br>**Self**: The sentiment is expressed with focus on himself/herself<br>e.g.:<br>• "This is driving me nutz:("<br>Mapping with categories of the OCC model [43]:<br>• Action of 'self' agent {approving/disapproving, pride/shame}<br>• Action of 'self' agent + Well being {gratification/remorse}<br>  Consequences of events on self {joy/distress, hope/fear, relief/disappointment, satisfaction/fear-confirmed}<br>**Other**: The sentiment is expressed towards the interlocutor<br>e.g.: "If anyone could help me that would be excellent! Thank you"<br>Mapping with categories of the OCC model [43]:<br>• Action of 'other' agents {approving/disapproving, admiration/reproach}<br>• Action of 'other' agents + Well being {gratitude/anger}<br>• Consequences of events on others {happy-for/resentment, gloating/pity} |
| Misclassification | **Sentiment Misclassification:** the detected sentiment does not match the true sentiment. Possible cases are humour (i.e. irony or sarcasm), linguistic features of politeness devices or misclassification of the domain lexicon,<br>e.g.:<br>- "Thanks in advance" (politeness); "Ignorance is bliss" (sarcasm); "I am missing a parenthesis but I don't know where" (neutral but annotated as negative by SentiStrength) |

### 4.2. Qualitative analysis of emotions

To further our understanding of the relationship between user *reputation* and *affect* in Stack Overflow, we perform a qualitative analysis of emotions identified in a subset of posts written by developers belonging to the four reputation categories. Consistently with previous research on emotion awareness in software engineering, we operationalize *affect* in terms of sentiment polarity using a state-of-the-art tool (i.e., SentiStrength) to capture the overall orientation of a text fragment as positive, negative, or neutral. However, sentiment polarity is not able to distinguish the specific attitude (i.e., 'friendly' vs. 'hostile') or emotion (i.e., 'joy' or 'anger') being expressed by the author of a post, nor to recognize the object of the sentiment being expressed in a text [21,42]. Thus, one of the authors performed content analysis on the questions conveying either positive or negative sentiment, according to the scores issued by SentiStrength. Overall, 240 posts were analyzed (120 with positive sentiment and 120 with negative sentiment) by randomly selecting the same number of questions for each reputation category.

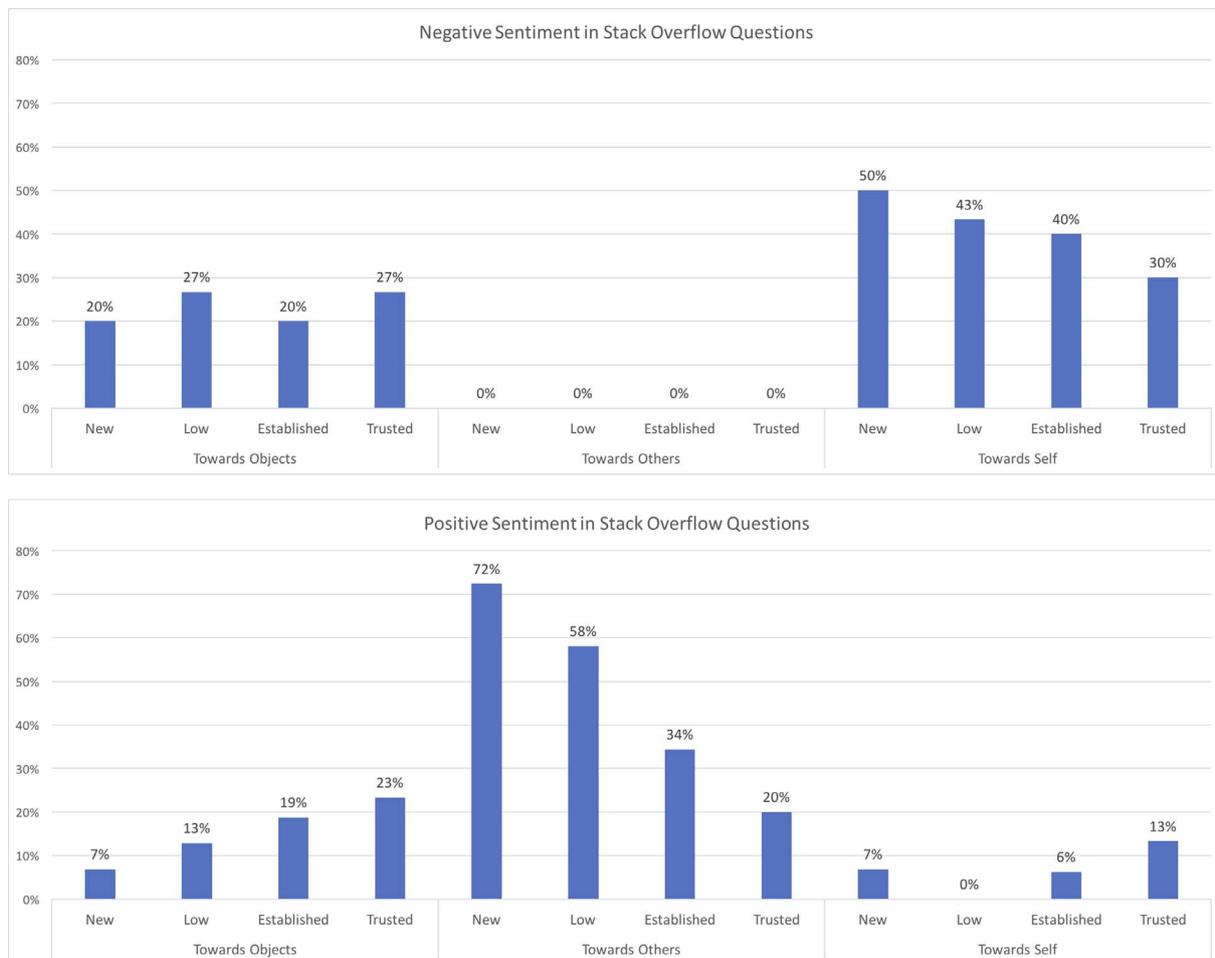

**Fig. 3.** Negative and positive sentiments in Stack Overflow questions.

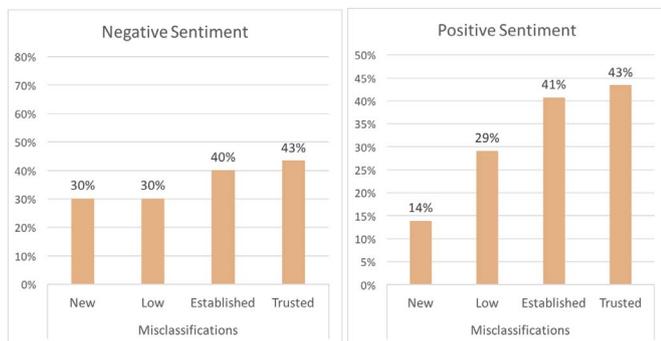

**Fig. 4.** Misclassification of sentiment polarity in Stack Overflow questions.

Content analysis involves the classification of data according to a finite set of possible categories. Thus, consistently with previous research [21], we defined coding labels (see Table 8) to capture:

- the *sentiment direction*, i.e., for those correctly classified, the target of the positive or negative sentiment expressed. According to the Ortony, Clore and Collins (OCC) model [43], emotions can be a reaction that focuses on *self* (as in the case of fear or joy), on the *other* agent (as in the case of approving/disapproving), or an attitude towards some properties of an *object* (as in the case of like/dislike).
- *misclassification*, i.e., whether we observe a mismatch between the sentiment detected by SentiStrength and the true sentiment conveyed by the question.

To check for biases, the coding of the questions was reviewed by another author. Then, disagreements were resolved iteratively through discussion. The final results of the coding are reported in Figs. 3 and 4 for both negative and positive sentiment. In particular, Fig. 3 depicts the distribution of emotion-direction codes for non-misclassified cases. Conversely, in Fig. 4 we report the proportion of misclassifications for each reputation category.

For both, negative and positive sentiment, we observe that the higher the user reputation, the more the misclassified cases. By analyzing the content of misclassified questions, we observe that questions posted by *Established* and *Trusted* users are richer in technical details and jargon, which is known to cause misclassifications when using sentiment-analysis tools that are not trained specifically for the software engineering domain [29,30,42].

As for the direction of negative sentiment, we found no cases of users showing a negative attitude towards community members (*other*). On the contrary, negative emotions are expressed by developers towards either *self* or an *object* (e.g. a programming language, a tool, a bug or a code snippet in the question). Negative emotions towards *self* are shown by *New* and *Low* reputation users, expressing frustration (as in '*I am totally lost on how it's supposed to work*', or '*A better way has to exist!*') or confusion (as in '*I am really confused with this question!*'). On the contrary, more expert users show less tendency to express such emotions. As for emotions towards *object*, we observe no differences with respect to the user reputation categories.

As for the direction of positive sentiment, we found that Stack Overflow users convey positive sentiment toward *others* in their questions, mainly to pay gratitude forward to potential helpers. However, as the reputation increases, we observe a decrease in the percentage of positive emotions towards *others*.

### 4.3. User survey

We arranged an online survey and then reached out to Stack Overflow users by advertising it on Twitter, Facebook, LinkedIn, and Research Gate. The survey (see Appendix A) included multiple choice, Likert scale, and open-ended questions. Since the survey was intended to provide support to the quantitative findings, we formulated one question for each of the predictors in our framework, according to the results of the logistic regression (see Tables 4 and 7). As for the open-ended responses, we reviewed them to identify common themes.

We received 46 responses in total, but 3 were removed because empty or containing only answers related to demographic. A breakdown of valid survey respondents ($n = 43$) is available in Appendix B. They are almost entirely males (38 vs 3 females), 30 years old on average (median 30; range 19–52), coming mostly from Italy (20), US (5), and Brazil (5) (see Fig. 5a). Besides, participants are evenly distributed regarding the field of occupation (see Fig. 5b). The survey participants also reported using Stack Overflow for 3.8 years on average (median 4; range 1–8; see Fig. 5c), mostly for asking questions (58.5%) (see Fig. 5d). Finally, albeit almost a half of the participants (19) opted not to answer this question, the breakdown of their reported reputation scores (mean ∼ 1944; median 133; range 0–12,000) shows that most of those who answered fall in the *Low reputation* category (see Fig. 5e).

In the following, we report the responses to the questions, grouped by success factor (see Fig. 6).

**Affect**. With respect to the emotional style of questions, the three Likert-scale questions show, clearly and consistently, that respondents think that questions written using a neutral style have better chances of success.

Overall, consistent results emerged from the analysis of the open question. The general opinion expressed by respondents is that "*Stack Overflow does not need emotions*" [P41] and that users should keep them out of questions "*so [the community] does not need to edit them*" [P2]. Consistently, respondent P6 noted that it is important to "*avoid being arrogant or urging for help, as Stack Overflow users answers for free.*" Besides, respondents noted that "*rude questions get rude [answers]*" [P26] and that "*if emotions take over the clarity of the question, this could probably lead to unsuccessful answers*" [P29]. On the contrary, respondent P17 reported an opposed perception that "*provocatively stated questions get more attention and more answers as a result.*" Finally, respondent P43 noted that typically strong emotions are expressed in response to "*questions that are difficult to understand (incomplete description of the situation, incomplete description of the task) or whose root issue is a novice developer trying to tackle a complex task and getting tangled in irrelevant details.*"

**Time**. Responses to the Likert-scale questions concerning time revealed that survey respondents have not a clear perception of 'when' posting a question is more effective, as shown by the large number of neutral responses. The analysis of the open-ended question related to time confirmed this finding about the effect of the day of week. For example, respondent P4 reported that "*maybe during weekends it may be possible to receive fewer visits for your questions.*" Instead, another respondent answered saying that "*people may be busy during workdays and not use Stack Overflow (if they don't need to)*" [P26].

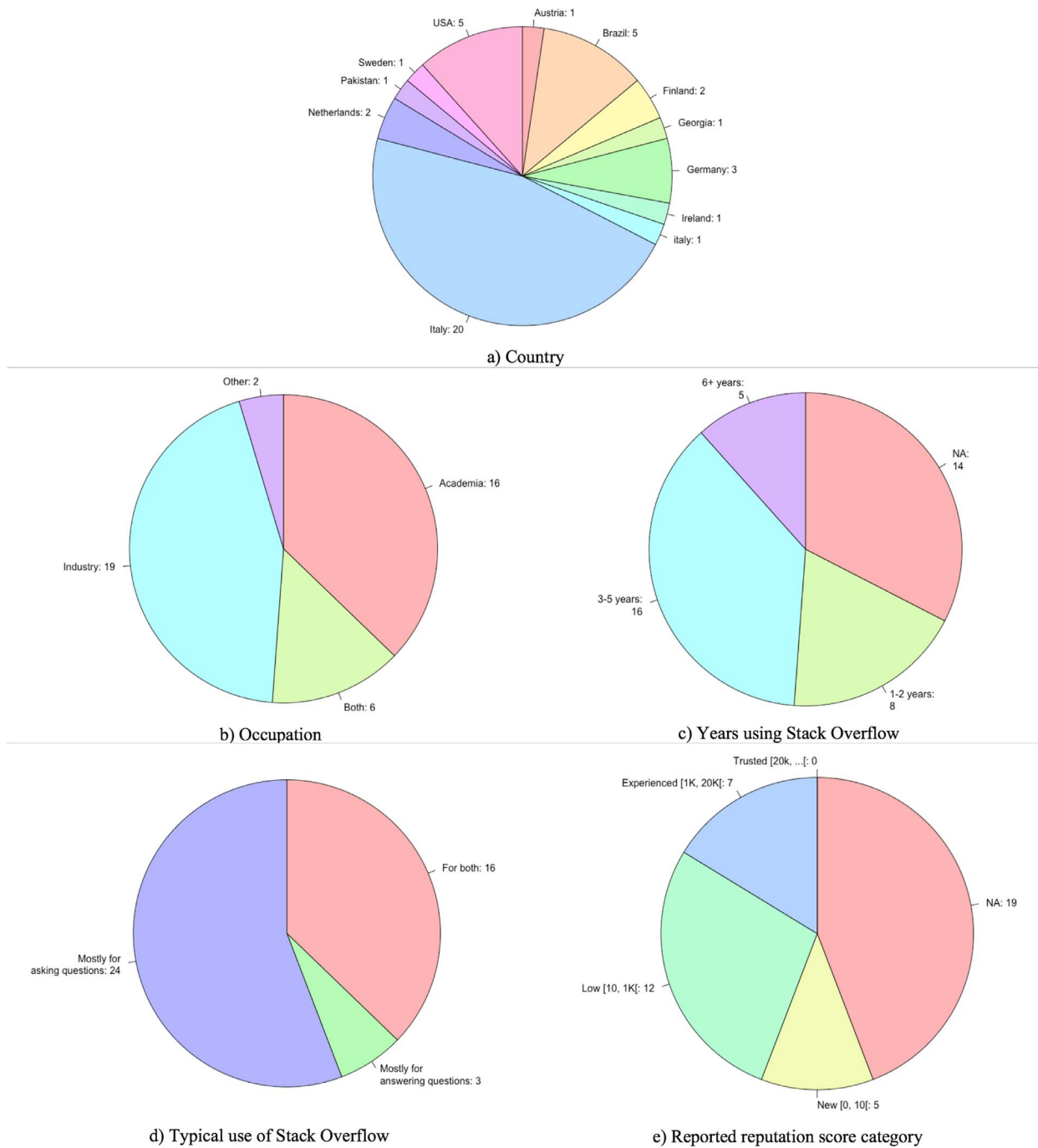

**Fig. 5.** Overview of survey respondents' demographic (*n* = 43).

**Presentation quality**. With respect to the effect of adding code snippets to questions, almost every respondent (95%) agreed that it would increase the chances of success. Besides, most of them agreed that labeling a question with multiple tags or including URLs to external resources increases its chances of getting a successful answer (62% and 49%, respectively), whereas questions that are abundant in uppercase characters have fewer chances of success (54%). Finally, no clear opinion emerges regarding the effect of writing concise questions with a short title.

According to the answers provided to the open-ended question, many respondents agreed that "*well-formatted and clear*" [P13, P34, P43] questions that "*show effort in describing the problem in a meaningful, technical way*" [P11, P23, P40] while also "*avoiding language errors*" [P4] are more likely to succeed. Other participants suggested also that using popular tags (i.e., asking questions relevant to a broader audience) [P22, P24] and choosing a "*wording close to popular Google searches*" [P28] can increase the chance of success.

Others, instead, commented that including code snippets in both question and answers is paramount. In fact, respondent [P33] noted that he does not mind about presentation style or correctness, as he immediately searches for a code snippet to focus on whether he is asking a question or looking for information. Consistently, respondent P24 noted that "*questions that include code snippets in a live-editor (e.g., JFiddle for JavaScript) have a higher chance of getting a successful answer.*" Finally, user P21 noted that all these features are relevant because what is relevant is "*totally dependent on the nature of the question. Some questions are best asked with code examples. But some [others] (e.g., about concepts like compiler optimizations and algorithms) are best described not with code, but with pictures.*"

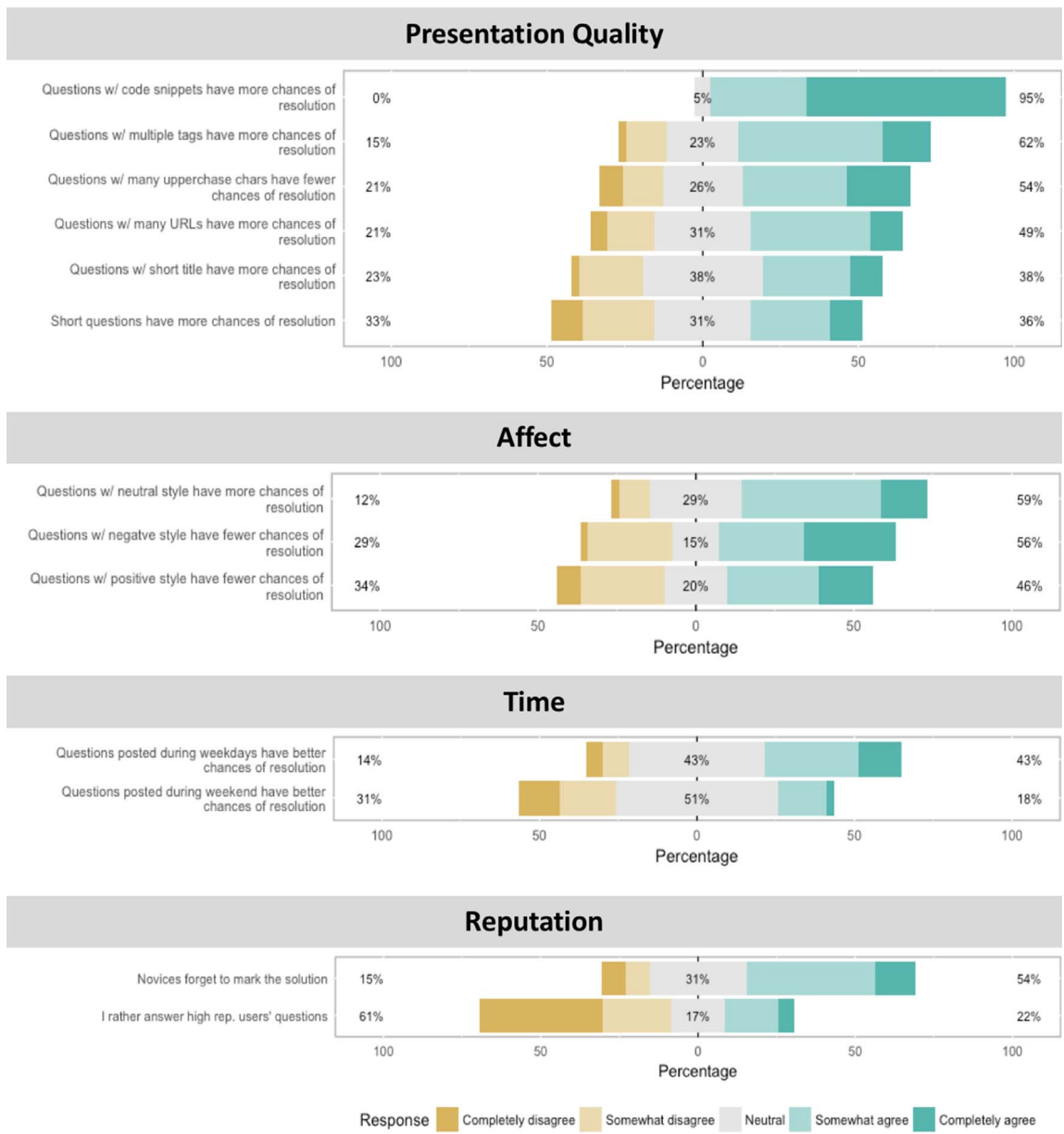

**Fig. 6.** Responses to Likert-scale questions, arranged by success factor (*n* = 43).

**Reputation**. When asked about novice users forgetting to mark the accepted solution, more than a half of the respondents (54%) agreed. Instead, most of the respondents (61%) disagree with the idea of rather answering questions posted by users with high reputation.

Regarding the open question, a few respondents underlined that reputation system in Stack Overflow favorites a 'rich get richer' effect.

In fact, users tend to think that if one's reputation is high then "*the answer is [received] from an expert, and this leads to [selecting it as] the successful answer*" [P29]. As such, user reputation in Stack Overflow "*reinforces itself – popularity means [one] gets more exposure, which gives them more of a chance to become popular*" [P17].

**Additional comments**. Albeit not specifically related to success, most of the responses to the last open-ended question "*Do you have any additional comments that you want to share with us?*" focus on the 'toxicity' of the site. One respondent reported that "*people on Stack Overflow can be extremely rude*" and, for that reason, the site is the very last resort because he does not "*really want to deal with [their] arrogant comments*" [P26]. Consistently, another participant answered that he prefers to "*access Stack Overflow [through Google searches] without a user account*" [P23]. Finally, one participant was specifically concerned about rudeness towards novice users, and students in particular, who often "*get*

**Table 9**
Summary of the validated set of guidelines for writing successful questions.

| Source | Guideline | Success factor | Supported? |
| --- | --- | --- | --- |
| Skeet [57], SO Help Center [58], Kucuktunc et al. [28], Bazelli et al. [6] | *Write questions using a neutral emotional style* | Affect | Yes |
| Skeet [57], Asaduzzaman et al. [4], Duijn et al. [17], Treude et al. [64] | *Provide sample code and data* | Presentation quality | Yes |
| Skeet [57] | *Use capital letters where appropriate* | Presentation quality | Yes |
| Skeet [57], | *Be concise* | Presentation quality | Yes |
| Skeet [57] | *Use short, descriptive question titles* | Presentation quality | No |
| Skeet [57] | *Provide context through tags* | Presentation quality | No |
| Ponzanelli et al. [49] (partially) | *Provide context through URLs* | Presentation quality | No |
| Bosu et al. [7] | *Be aware of low-efficiency hours* | Time | Yes |

*flamed in this forum*" and, therefore, are "*unlikely to return, ever*" [P28].

## 5. Discussion

We discuss the findings of the current study and their main implications both for practitioners and researchers, respectively in Section 5.1 and 5.2.

### 5.1. Practical implications: how to write a good question

In this section, we discuss the set of guidelines for writing good questions listed in Table 9. These guidelines, for which we sought supporting evidence through our study, are discussed in the following, grouped by success factor (i.e., *Affect, Presentation quality*, and *Time*). Instead, the discussion of the effects of *Reputation* – i.e., the control (non-actionable) factor in our framework of analysis – crosscuts the discussion of the previous three success factors.

**Affect**. *Write questions using a neutral emotional style*. Regarding the affective aspect of questions, we can say that the invitation to adopt a neutral style has been by now accepted by the Stack Overflow community. In fact, according to the results of our survey, most of the respondents agreed that adopting a neutral communication style when writing questions may increase their chance of success (see Fig. 6 in Section 4.3). Moreover, none of the questions analyzed contains negative emotions (i.e., rudeness) directed towards others. In particular, the results of the qualitative analysis (see Fig. 3) show that, rather than towards the potential helpers, the negative sentiment expressed in questions is actually detected because of a negative attitude towards the asker herself or the technical issue. It is the case of negative emotions, such as frustration for not being able to solve a problem or negative opinions about something that does not work as expected. Similarly, positive sentiment lexicon in questions is mainly used for paying gratitude in advance to potential helpers. This might be interpreted as a consequence of the successful moderation accomplished by the Stack Overflow community. In fact, Stack Overflow relies on a social reputation system designed to both incentivize contributions and enable the assessment of users' trustworthiness through the upvotes and downvotes given to both questions and answers. However, previous studies suggested that negative and hostile attitude is likely to be expressed in comments rather than in questions because comments are excluded by the reputation system (i.e., cannot be voted) and thus, they are seen as a free zone [9,42]. Consistently, many survey respondents answered the final open question (*Do you have any additional comments that you want to share with us?*) complaining against the arrogance and rudeness of some Stack Overflow users' comments. A few respondents explicitly described the harsh tone of comments as a possible deal breaker for many. The inclusion of comments in the reputation mechanism has been under discussion among users at Meta SO, yet the implementation of the feature has been declined for now [34]. On the contrary, these results suggest that Stack Overflow should instead implement comment moderation sooner rather than later.

*Expressing emotions is associated with lower probability of success.* Regarding the sentiment in the question body, we observe that expressing negative emotions is associated with the low probability of success (see Table 4). This is further confirmed by the results of the logistic regression analysis performed for each reputation category (see Table 7). Furthermore, we observe a stronger effect size for *Trusted* users. In the scenario of a technical Q&A site, being able to distinguish attitudes towards the reader (friendly vs. hostile) from judgments about the technical issue could be helpful for the community of moderators [21]. The results of the content analysis show that a hostile attitude towards the interlocutors never occurs in questions. On the contrary, we observe a tendency for *New* and *Low Reputation* users to express frustration and confusion in their questions, which is the main reason for detection on negative sentiment in Stack Overflow questions.

Finally, we found that expressing positive sentiment within the question body (mainly gratitude towards potential helpers), is associated with a lower probability of success. However, the effect size is small and the correlation of the positive sentiment with success of the question is statistically significant only for *Low* and *Established* users. Albeit this finding may seem counterintuitive – one may assume that having a positive attitude towards the community members would induce people to help – it is consistent with the findings of Bazelli et al. [6] who observed that questions containing lexical cues of extroversion are downvoted more often. Furthermore, it is consistent with Skeet's suggestion of avoiding positive lexicon for expressing greetings or paying gratitude forward, which could result in extra effort for potential helpers when reading the question [57].

Overall, this evidence confirms that adopting a neutral writing style improves the perception of quality not only for answers [28] but also for questions.

**Presentation quality**. *Provide sample code and data*. We confirm the importance of providing sample code in questions for the sake of clarity and completeness of the information provided. This finding is also consistent with both the community guidelines suggested by Skeet and the evidence provided by previous research [4,17,64]. The importance of attaching code snippets to questions is clear to community members since 95% of survey respondents strongly agreed that it would enhance the chance of receiving successful answers. Besides, they suggested the importance of using live code editors, which would enable community users to fiddle directly with the code samples provided by askers.

*Use capital letters where appropriate and be concise.* We are also able to confirm that making inappropriate use of capital letters impairs readability and might indicate unawareness of the netiquette, as their misuse is usually perceived as yelling. Besides, we confirm that too long questions should be avoided since people do not like to read too much. This is especially true for top-reputation users (*Trusted* users) for which we observe a stronger negative correlation between question length and success (OR = 0.31 for *Long* questions, see Table 7). Thus, the results of our study provide empirical support to these informal recommendations provided by the Stack Overflow community, suggesting a potential asker to focus on conciseness and completeness when writing a question. Still, it is interesting to note that, based on the survey analysis, the Stack Overflow community members are not aware of the effectiveness of writing short questions (see Fig. 6). Hence, Stack Overflow should take actions to increase the visibility of these presentation quality guidelines and possibly check for violations in the text editor automatically.

**Table 10**
Distribution of edited vs unedited questions by reputation category.

| Reputation | Unedited | Edited |
| --- | --- | --- |
| NEW Users (Reputation Score < 10) | 45% | 55% |
| LOW Reputation Users (Score in [10, 10K[) | 52% | 48% |
| ESTABLISHED Users (Score in [10 K, 20K[) | 54% | 46% |
| TRUSTED Users (Score > = 20 K) | 57% | 43% |

Conversely, we are not able to confirm the other guidelines about using short titles that capture the theme of the question (*Use short, descriptive question titles*), and adding multiple tags to provide context for a question (*Provide context through tags* and *Provide context through URLs*), for which we observe contrasting evidence in significance and odds ratio values depending on the user reputation category (see Table 7). Interestingly, this evidence contrasts not only with the community guidelines by Skeet but also with the perception of Stack Overflow users, who mostly agreed with the usefulness of adding tags and URLs to increase the chance of success of questions (see Fig. 6).

**Time**. *Be aware of low-efficiency hours and days*. In line with previous research [7], we confirm that questions posted during some time slices are more successful than others. In particular, we confirm that the most successful time slices are evening and night in GMT zone, corresponding to American working time. However, we could not find support regarding previous findings suggesting that posting questions to Stack Overflow during weekend correlates to a higher chance of eliciting successful answers; albeit a positive correlation was observed, the effect size was negligible (OR = 1.10).

Despite the impact of the time factor might vary in the future as the community demography changes, users should be aware that low-efficiency hours exist and that the Stack Overflow community might be more reactive during evening and night GMT hours. Instead, the results of the survey show that most of the users ignore the existence of such 'more successful' time slices.

### 5.2. Theoretical implications

*Theoretical framework of success factors*. Other than practical implications for developers, the results of our studies also suggest theoretical insights for researchers. In particular, our analysis was informed by a theoretical framework that we built by defining actionable and non-actionable factors for the success of Stack Overflow questions (see Section 2.2). We operationalize our framework using a set of metrics (see Section 2.3), which we treat as predictors in a logistic regression framework. Results of our studies (see Table 4 in Section 4) show that our model significantly differs from the null model (intercept-only model), albeit with a low AUC = 0.65. A possible explanation is that the Stack Overflow community is vast and heterogeneous and we might have overlooked other non-actionable factors of success. Among those, availability of experts has been correlated with the probability of success of a question [7]. Similarly, the popularity of a topic and the degree of maturity of a technology might affect the probability of success of a question.

*Theoretical models of affect*. As for emotions, we operationalize affective states in terms of polarity (positive, negative, or neutral). However, the qualitative analysis of emotions and their direction, performed on a sample of 240 questions (see Section 4.2), advocate in favor of a more fine-grained approach for operationalizing affective states [21]. For example, as for the negative sentiment direction, we found no cases of users showing a negative attitude towards the community (i.e., towards *other*). Conversely, we found *self*-directed emotions, such as frustration and sadness, which do not involve a negative attitude towards the interlocutor. This is an evidence that cannot be obtained by simply looking at the negative polarity of sentiment expressed in the question body.

In contrast with the evidence provided by our qualitative analysis, previous studies report that negative attitude is actually expressed by Stack Overflow users. However, hostile behavior is detected mainly in comments, probably because it is a reputation-free zone [42]. In fact, Stack Overflow does not implement a downvote feature for comments and comment upvotes do not contribute to reputation. However, Asaduzzaman et al. [4] suggest that strong negative sentiment expressed by askers in follow-up discussions may discourage user participation in Stack Overflow. In a further replication of this study, researchers might consider analyzing asker behavior also in the *question nurturing* phase, that is the follow-up discussion originated by the original question.

*Asker reputation as the main predictor of success for questions*. Finally, we observe that asker *reputation* is the main predictor of success for Stack Overflow questions. Based on empirical evidence provided in this study, one possible reason for this is the unawareness of community rules by *New* users, who tend not to mark useful answers as a solution in 38% of cases (see the results of the qualitative analysis of unsuccessful questions by *New* users in Table 5). Another explanation might be related to the higher propensity of the community members to help information seekers with high reputation. While this explanation would be consistent with the results from previous research [2], we found no supporting evidence in our study. This suggests directions for future investigation on the impact of asker reputation on the success of technical requests through controlled experiments.

## 6. Limitations

One of the key issues in empirical software engineering is evaluating the validity of results [65]. In the following, we discuss threats to the validity of our findings.

*Threats to external validity*. External validity of a study concerns to its representativeness and to the ability to generalize the conclusions beyond the scope of the study itself. In this paper, we present the results of a study specifically focused on Stack Overflow. Therefore, our methodology could produce different results if applied to other technical Q&A sites. We acknowledge this limitation and we intend to address it in future replications.

*Threats to internal validity*. Internal validity influences the conclusions about a possible causal relationship between the treatment and the result of a study. Analogously to previous research [11,64], we consider as successful a question with an accepted answer. However, questions may remain without an accepted answer for diverse reasons, not necessarily associated with the dissatisfaction of the information seeker. For example, an inexperienced asker might forget to mark as accepted the best answers because indifferent or unaware of the community rules, as confirmed by the gratitude expressed in many asker's comments as a reply to helpful answers [9].

Another threat to internal validity concerns the construction of the dataset, which we obtained by removing all edited questions. We acknowledge that our approach may have potentially filtered out some questions useful to our research purpose. To assess the impact of this preprocessing step, we computed the edited vs. unedited question distribution in the current Stack Overflow data dump.[7] Results are reported in the Table 10 and show that questions posted by higher reputation members (i.e., established and trusted) are less likely to be edited. However, the percentages for each reputation category are comparable, thus mitigating concerns about this limitation.

---
[7] https://data.stackexchange.com/stackoverflow/query/new

Furthermore, because Stack Overflow does not provide the history of changes but only the final form of edited questions, filtering-out edited questions became a necessary preprocessing step for our goal of understanding the text-related factors that increase the chances of being resolved – i.e., what a question looked like *when* the successful answer was posted. Hence, we note that this issue will be otherwise addressable only when Stack Overflow decides to provide the history of edits in their future data dumps.

*Construct validity.* Threats to construct validity concern the degree of accuracy to which the variables defined in a study measure the constructs of interest. As for the measures defined in our analysis framework, most of the variables have been operationalized as ordinal (e.g., *Body Length*). We acknowledge that different choices of category thresholds (e.g., *Body Length* in [0, 90[characters as *Short*) may generate different results. Nonetheless, to control for this threat and avoid choosing arbitrarily the range of values associated with each category, in the study we opted for relying on the k-means clustering algorithm from the *arules* R package. Instead, regarding the *Asker Reputation* variable, we decided to comply with the same reputation categories (i.e., *New, Low, Established*, and *Trusted*) and associated thresholds defined by Stack Overflow, being it the actual platform under study.

Consistently with previous studies on affect in empirical software engineering, we have measured the sentiment polarity of a text, that is its positive or negative overall orientation. However, the results of our content analysis highlight that affectively-loaded lexicon is used for expressing a wide range of affective states, thus calling for further research on emotion detection in software engineering, which still represents an open research challenge [29,39,42].

Another threat to construct validity regards the tool used for sentiment detection. Although we used a state-of-the-art tool that has been validated using a data set obtained from six different social sites [63], the problem of domain-dependence still exists [5,20], especially for technical sites such as Stack Overflow. For example, Jongeling et al. [29,30] conducted a classification study of seven datasets from technical websites (i.e., issue trackers and Stack Overflow questions) using different sentiment analysis tools and observed that the disagreement between the tools leads to diverging conclusions and lack of replicability. To mitigate this threat, we validated SentiStrength on a gold standard of 400 manually annotated questions, answers, and comments from Stack Overflow. Each document was manually annotated by each of the three authors, which indicated its sentiment polarity, with a possible value in *{positive, negative, neutral}*. When performing the annotation, positive polarity was indicated whenever a positive emotion, such as joy or gratitude, was detected. Conversely, negative polarity was indicated when a negative emotion, such as anger or sadness, was identified. The neutral label was used to indicate the absence of emotion. The final polarity label for each item was assigned using majority voting (two out of three indicating the same label), thus resulting in a balanced dataset of 344 items for which the majority voting was reached. The dataset contains 35% negative items, 38% positive items, and 27% neutral items, which we release for public use for research purposes.[8] We observed a substantial agreement among raters with average Kappa = 0.74. The performance of SentiStrength on the gold set (Accuracy = 0.80, Precision = 0.77, Recall = 0.79, F-measure = 0.78) are in line with state of the art on sentiment analysis on social media [40]. This evidence mitigates the envisaged threat to construct validity for this study. However, the misclassification rate of content analysis in presence of technical jargon or error messages indicates a problem with misclassification of neutral cases as negative (see Section 4.2). In particular, the high number of false positives for negative sentiment annotation might be due to the goal of a technical Q&A site, which is explicitly designed for people seeking help. Therefore, discussions tend to be skewed towards negative polarity, without any real intention to show an affective state, as in the following example, for which a negative score is inaccurately issued by SentiStrength: "*I have a trouble […]. Please, explain what is wrong.*" The empirical evidence provided in this study supports the need for SE-specific approaches to sentiment analysis, involving tuning of existing sentiment lexicons and tools. Recently, a customized version of SentiStrength has been developed to support sentiment analysis in software engineering [51] as well as an emotion arousal lexicon [32] and emotion classifiers [10,21] specifically designed for mining affect in software engineering texts. At the time of the analysis, these resources were unavailable. As such, future replications will involve the use of such tools to further validate the findings of the current study.

Finally, in this study, we chose to exclude non-actionable factors that have been shown to predict the success of requests, such as expert availability [7]. Although consistent with our goal of building a set of practical, evidence-based guidelines for writing questions, this choice might be a limitation for building a general model of success predictors in Q&A sites where other variables, such as the topic or the maturity of a technology, may play a role in the success of questions.

## 7. Conclusions

In this paper, we presented an empirical investigation on the impact of *affect, presentation quality*, and *time* on the success of Stack Overflow questions. We provided evidence-based guidelines that information seekers can follow to increase the chance of getting help. As for *presentation quality*, adding code snippets to the body of a question is strongly recommended. Users are also advised to be concise and avoid unnecessary use of uppercase characters. Furthermore, with respect to *time*, information seekers should be aware that low and high-efficiency hours exist.

As for the role of *affect*, we empirically confirm community guidelines that suggest avoiding rudeness in question writing. We also observed that a neutral writing style is associated with a higher probability of success. This is in line with evidence provided by previous research about online Q&A sites, thus indicating that expression of sentiment, regardless of the polarity, might be detrimental to success.

Content analysis has revealed open challenges for sentiment detection in software engineering and inspires directions for future work. Measuring the polarity of a text might be complemented with information about emotion direction to enable moderation when a negative attitude towards the interlocutor is detected. On the contrary, a negative attitude towards oneself or problem sources might indicate the need for prioritization in addressing help requests. Being able to detect such situations might be relevant for recommending controversial discussion threads to experts and project leaders.

Finally, we underline the need for tuning state-of-the-art resources for polarity classification to overcome the limitations induced by domain-dependent use of a lexicon. New sentiment analysis tools specific to the software engineering discipline might be developed in order to distinguish accurately neutral sentences from emotionally loaded ones in technical discussions.

## Acknowledgment

This work is partially supported by the project 'EmoQuest - Investigating the Role of Emotions in Online Question & Answer Sites', funded by the Italian Ministry of Education, University and Research (MIUR) under the program "Scientific Independence of young Researchers" (SIR). The computational work has been executed on the IT resources made available by two projects, ReCaS and PRISMA, funded by MIUR under the program "PON R&C 2007–2013".

---

[8] The gold standard is available for download at: https://goo.gl/V6p33f

**Appendix A. – Survey**

## Demographic

Questions are *all optional*
Note: Below, an answer is said to be successful when it is marked by the asker as the accepted solution.

1. **Age**
   ______________________

2. **Country**
   ______________________

3. **What is your occupation?**
   *Check all that apply.*
   ☐ Academic (student, researcher, ...)
   ☐ Industry (developer, ...)
   ☐ Other:

**Gender**
*Mark only one oval.*
○ Male
○ Female
○ Prefer not to say
○ Other: ______________________

5. **You go on Stack Overflow...**
   *Mark only one oval.*
   ○ mostly for asking questions
   ○ mostly for answering questions
   ○ for both asking and answering questions

6. **Your Stack Overflow username**
   This is *voluntary* and it is *not a requirement* of the survey. This will give us some context for your responses. Leave it blank if you're uneasy about it!
   ______________________

7. **Years since you joined Stack Overflow**
   ______________________

8. **Your current Stack Overflow reputation score**
   ______________________

### Reputation

9. **I prefer to answer questions posted by users with high reputation.**
   *Mark only one oval.*

   |   | 1 | 2 | 3 | 4 | 5 |   |
   |---|---|---|---|---|---|---|
   | Completely disagree | ◯ | ◯ | ◯ | ◯ | ◯ | Completely agree |

10. **Novice users forget to mark the best received answer as the accepted solution.**
    *Mark only one oval.*

    |   | 1 | 2 | 3 | 4 | 5 |   |
    |---|---|---|---|---|---|---|
    | Completely disagree | ◯ | ◯ | ◯ | ◯ | ◯ | Completely agree |

11. **Is there anything else related to reputation that may influence the chances of getting a successful answer in Stack Overflow?**

    ______________________________________
    ______________________________________
    ______________________________________
    ______________________________________

### Time

12. **Questions posted during the weekend have more chances of getting a successful answer.**
    *Mark only one oval.*

    |   | 1 | 2 | 3 | 4 | 5 |   |
    |---|---|---|---|---|---|---|
    | Completely disagree | ◯ | ◯ | ◯ | ◯ | ◯ | Completely agree |

13. **Questions posted during working time have more chances of getting a successful answer.**
    *Mark only one oval.*

    |   | 1 | 2 | 3 | 4 | 5 |   |
    |---|---|---|---|---|---|---|
    | Completely disagree | ◯ | ◯ | ◯ | ◯ | ◯ | Completely agree |

14. **Is there anything else related to time that may influence the chances of getting a successful answer in Stack Overflow?**

   _______________________________________
   _______________________________________
   _______________________________________
   _______________________________________

## Emotional style

15. **Questions with a neutral emotional style have more chances of getting a successful answer.**
    *Mark only one oval.*

    |                     | 1 | 2 | 3 | 4 | 5 |                  |
    |---------------------|---|---|---|---|---|------------------|
    | Completely disagree | ◯ | ◯ | ◯ | ◯ | ◯ | Completely agree |

16. **Questions with a positive style (e.g., expressing gratitude in advance) have fewer chances of getting a successful answer.**
    *Mark only one oval.*

    |                     | 1 | 2 | 3 | 4 | 5 |                  |
    |---------------------|---|---|---|---|---|------------------|
    | Completely disagree | ◯ | ◯ | ◯ | ◯ | ◯ | Completely agree |

17. **Questions with a negative style (e.g., showing frustration) have fewer chances of getting a successful answer.**
    *Mark only one oval.*

    |                     | 1 | 2 | 3 | 4 | 5 |                  |
    |---------------------|---|---|---|---|---|------------------|
    | Completely disagree | ◯ | ◯ | ◯ | ◯ | ◯ | Completely agree |

18. **Is there anything else related to the emotional style of a question that may influence the chances of getting a successful answer in Stack Overflow?**

   _______________________________________
   _______________________________________
   _______________________________________
   _______________________________________

## Presentation quality

19. **Questions with code snippet(s) have more chances of getting a successful answer.**
    *Mark only one oval.*

    |  | 1 | 2 | 3 | 4 | 5 |  |
    |---|---|---|---|---|---|---|
    | Completely disagree | ◯ | ◯ | ◯ | ◯ | ◯ | Completely agree |

20. **Questions with many uppercase characters have fewer chances of getting a successful answer.**
    *Mark only one oval.*

    |  | 1 | 2 | 3 | 4 | 5 |  |
    |---|---|---|---|---|---|---|
    | Completely disagree | ◯ | ◯ | ◯ | ◯ | ◯ | Completely agree |

21. **Short questions have more chances of getting a successful answer.**
    *Mark only one oval.*

    |  | 1 | 2 | 3 | 4 | 5 |  |
    |---|---|---|---|---|---|---|
    | Completely disagree | ◯ | ◯ | ◯ | ◯ | ◯ | Completely agree |

22. **Questions with a short title have more chances of getting a successful answer.**
    *Mark only one oval.*

    |  | 1 | 2 | 3 | 4 | 5 |  |
    |---|---|---|---|---|---|---|
    | Completely disagree | ◯ | ◯ | ◯ | ◯ | ◯ | Completely agree |

23. **Questions with multiple tags have more chances of getting a successful answer.**
    *Mark only one oval.*

    |  | 1 | 2 | 3 | 4 | 5 |  |
    |---|---|---|---|---|---|---|
    | Completely disagree | ◯ | ◯ | ◯ | ◯ | ◯ | Completely agree |

24. **Questions with links to external resources (URLs) have more chances of getting a successful answer.**
    *Mark only one oval.*

    |  | 1 | 2 | 3 | 4 | 5 |  |
    |---|---|---|---|---|---|---|
    | Completely disagree | ◯ | ◯ | ◯ | ◯ | ◯ | Completely agree |

25. **Is there anything else related to the presentation style of a question that may influence the chances of getting a successful answer in Stack Overflow?**

   _______________________________________________
   _______________________________________________
   _______________________________________________
   _______________________________________________
   _______________________________________________

26. **Do you have any additional comments that you want to share with us?**

   _______________________________________________
   _______________________________________________
   _______________________________________________
   _______________________________________________
   _______________________________________________

**Would you like to be informed about the outcome of this study and potential publications?**

Enter a valid email below. This is *voluntary* and it is *not a requirement* of the survey.
We will only use the email provided to notify you of the results. We also promise to never disclose your address to anyone and to destroy any record of it after sending the notification.

27. **Your email address**

   _______________________________________________

Appendix B. – Breakdown of survey respondents

| Id | Age | Gender | Country | Occupation | SO experience (years) | SO reputation | Typical use of SO |
|----|-----|--------|---------|------------|----------------------|---------------|-------------------|
| P1 | 29 | Male | Austria | Academia | 4 | 9408 | mostly for answering questions |
| P2 | 47 | Male | Italy | Industry | 2 | 12,000 | mostly for answering questions |
| P3 | 30 | Male | Italy | Industry | 4 | 204 | mostly for asking questions |
| P4 | 35 | Male | Italy | Industry | 7 | 182 | mostly for asking questions |
| P5 | 29 | Male | Italy | Industry | 5 |  | mostly for asking questions |
| P6 | 32 | Male | Italy | Industry | 2 | 128 | mostly for asking questions |
| P7 | 29 | Male | Italy | Industry | 4 |  | for both asking and answering questions |
| P8 | 30 | Male | Italy | Industry | 5.5 | 50 | for both asking and answering questions |
| P9 | 29 | Male | Italy | Academia, Industry |  |  | mostly for asking questions |
| P10 | 29 | Male | Italy | Academia, Industry |  |  | mostly for asking questions |
| P11 | 31 | Male | Georgia | Industry | 3 | 12,000 | for both asking and answering questions |
| P12 | 33 | Male | Netherlands | Academia | 2 | 1 | for both asking and answering questions |
| P13 | 34 | Male | Italy | Industry |  |  | mostly for asking questions |
| P14 | 31 | Male | Italy | Industry | 3 | 223 | for both asking and answering questions |
| P15 | 25 | Male | Italy | Industry | 2 | 0 | mostly for asking questions |
| P16 | 25 | Male | Italy | Academia | 3 | 8 | mostly for asking questions |
| P17 | 52 | Male | Netherlands | Other | 8 | 5866 | for both asking and answering questions |

| P18 | 25 | Male | Italy | Industry | 3 | 15 | mostly for asking questions |
| P19 | 47 | Male | Finland | Academia | | | mostly for asking questions |
| P20 | 23 | Female | Brazil | Academia | 4 | 487 | mostly for asking questions |
| P21 | 27 | Male | USA | Academia | 4 | 48 | mostly for asking questions |
| P22 | 32 | Male | Sweden | Academia, Industry | 3 | 1 | for both asking and answering questions |
| P23 | 33 | Male | Pakistan | Industry | 6 | | mostly for asking questions |
| P24 | 29 | Male | Germany | Academia, Industry | | | for both asking and answering questions |
| P25 | 32 | Male | Germany | Academia | | | mostly for asking questions |
| P26 | 30 | Male | United States | Academia | 1.5 | 25 | mostly for asking questions |
| P27 | 25 | Male | USA | Academia, Industry | 7 | | for both asking and answering questions |
| P28 | 31 | Male | USA | Industry | 1 | 0 | for both asking and answering questions |
| P29 | 23 | Male | Italy | Academia | 2 | 1 | mostly for asking questions |
| P30 | 27 | Male | Germany | Academia | | | mostly for asking questions |
| P31 | 26 | Male | Brazil | Academia, Industry | | | mostly for asking questions |
| P32 | 22 | Male | Brazil | Academia | 4 | | mostly for asking questions |
| P33 | 37 | Male | Finland | Industry | | | mostly for asking questions |
| P34 | 30 | Male | Italy | Industry | 5 | 138 | for both asking and answering questions |
| P35 | 30 | Male | Italy | Academia | 2 | 1 | mostly for asking questions |
| P36 | 20 | Male | Brazil | Academia | | | mostly for asking questions |
| P37 | 19 | Male | Brazil | Academia | 5 | 220 | mostly for asking questions |
| P38 | 39 | Female | USA | Academia | 2 | | mostly for answering questions |
| P39 | 33 | Female | Ireland | Industry | 4 | 108 | for both asking and answering questions |
| P40 | 29 | Male | Italy | Industry | | 1573 | for both asking and answering questions |
| P41 | 32 | Male | Italy | Industry | Idk | 2260 | for both asking and answering questions |
| P42 | 35 | Male | Italy | Academia | | | for both asking and answering questions |
| P43 | 40 | Male | Italy | Industry | 5 | 5600 | for both asking and answering questions |

**References**

[1] R. Abdalkareem, E. Shihab, J. Rilling, What do developers use the crowd for? A study using stack overflow, IEEE Softw. 34 ((March)2) (2017) 53–60.
[2] T. Althoff, C. Danescu-Niculescu-Mizil, D. Jurafsky, How to ask for a favor: a case study on the success of altruistic requests, Proceeding of the 8th International AAAI Conference on Weblogs and Social Media (ICWSM 2014), 2014.
[3] A. Anderson, D. Huttenlocher, J. Kleinberg, J. Leskovec, Discovering value from community activity on focused question answering sites: a case study of stack overflow, Proc. of 18th ACM SIGKDD Int'l Conference on Knowledge Discovery and Data mining (KDD '12), ACM, 2012, pp. 850–858.
[4] M. Asaduzzaman, A.S Mashiyat, C.K. Roy, K.A. Schneider, Answering questions about unanswered questions of stack overflow, Proc. of the 10th IEEE Working Conf. on Mining Software Repositories (MSR 2013), 97 2013, p. 100.
[5] P. Basile, V. Basile, M. Nissim, N. Novielli, Deep tweets: from entity linking to sentiment analysis, Proc. of the 2nd Italian Conference on Computational Linguistics (CLIC'15), 2015.
[6] B. Bazelli, A. Hindle, E. Stroulia, On the personality traits of stack overflow users, Proc. of the 2013 IEEE Int'l Conference on Software Maintenance (ICSM '13), USA, IEEE Computer Society, 2013, pp. 460–463.
[7] A. Bosu, C.S. Corley, D. Heaton, D. Chatterji, D.J. Carver, N.A. Kraft, Building reputation in stack overflow: an empirical investigation, Proc. of the 10th IEEE Working Conf. on Mining Software Repositories (MSR 2013), 89 2013, p. 92.
[8] F. Calefato, F. Lanubile, M.R. Merolla, N. Novielli, Success factors for effective knowledge sharing in community-based question-answering, Proc. 10th International Forum on Knowledge Asset Dynamics (IFKAD'15), 2015.
[9] F. Calefato, F. Lanubile, M.C. Marasciulo, N. Novielli, MSR challenge: mining successful answers in stack overflow, Proc. of 12th IEEE Working Conf. on Mining Software Repositories (MSR 2015), 2015.
[10] F. Calefato, F. Lanubile, N. Novielli, EmoTxt: a toolkit for emotion recognition from text, Proceedings of the 7th Affective Computing and Intelligent Interaction (ACII'17), San Antonio, TX, USA, 2017 Oct. 23-26 (in press).
[11] F. Calefato, F. Lanubile, N. Novielli, Moving to stack overflow: best-answer prediction in legacy developer forums, Proceedings of the 10th ACM/IEEE International Symposium on Empirical Software Engineering and Measurement (ESEM '16), ACM, 2016.
[12] V. Carofiglio, F. de Rosis, N. Novielli, Cognitive Emotion Modeling in Natural Language Communication, Springer, LondonLondon, 2009, pp. 23–44.
[13] M. Cataldo, J.D. Herbsleb, K.M. Carley, Socio-technical congruence: a framework for assessing the impact of technical and work dependencies on software development productivity, Proceedings of the Second ACM-IEEE International Symposium on Empirical Software Engineering and Measurement (ESEM '08), 2008, pp. 2–11.
[14] Collinearity and Stepwise VIF Selection - https://beckmw.wordpress.com/2013/02/05/collinearity-and-stepwise-vif-selection (Last accessed: June 2017).
[15] L. Dabbish, C. Stuart, J. Tsay, J.D. Herbsleb, Social coding in GitHub: transparency and collaboration in an open software repository, Proceedings of the ACM 2012 Conference on Computer Supported Cooperative Work (CSCW '12), 2012, pp. 1277–1286.
[16] G. Destefanis, M. Ortu, S. Counsell, S. Swift, M Marchesi, R Tonelli, Software development: do good manners matter? Peer J. Comput. Sci. 2 (2016) e73 https://doi.org/10.7717/peerj-cs.73.
[17] M. Duijn, A. Kučera, A. Bacchelli, Quality questions need quality code: classifying code fragments on stack overflow, Proceedings of the 12th Working Conference on Mining Software Repositories (MSR '15), 2015, pp. 410–413.
[18] P. Ekman, Basic emotions, Handbook of Cognition and Emotion, John Wiley & Sons Ltd, 1999.
[19] S. Easterbrook, J. Singer, M.-A. Storey, D. Damian, Selecting empirical methods for software engineering research, Guide to Advanced Empirical Software Engineering, Springer, 2008, pp. 285–311.
[20] M. Gamon, A. Aue, S. Corston-Oliver, E. Ringger, Pulse: mining customer opinions from free text, Lecture Notes in Computer Science 3646 (2005) 121–132.
[21] D. Gachechiladze, F. Lanubile, N. Novielli, A. Serebrenik, Anger and its direction in collaborative software development, Proc. of ICSE 2017, 2017, NIER Track.
[22] D. Garcia, M.S. Zanetti, F. Schweitzer, The role of emotions in contributors' activity: a case study on the Gentoo community, International Conference on Cloud and Green Computing, 2013, pp. 410–417.
[23] E. Guzman, D. Azócar, Y. Li, Sentiment analysis of commit comments in GitHub: an empirical study, Proc. of the 11th Working Conf. on Mining Software Repositories (MSR 2014), New York, NY, USA, ACM, 2014, pp. 352–355.
[24] E. Guzman, B. Bruegge, Towards emotional awareness in software development teams, Proc. of the 2013 9th Joint Meeting on Foundations of Software Engineering (ESEC/FSE 2013), New York, USA, 2013, pp. 671–674.
[25] E. Guzman, W. Maalej, How do users like this feature? A fine grained sentiment analysis of app reviews, 22nd International Conference on Requirements Engineering (RE'14), IEEE, 2014, pp. 153–162.
[26] D. Graziotin, X. Wang, P. Abrahamsson, Do feelings matter? On the correlation of affects and the self-assessed productivity in software engineering, J. Softw. 27 (7) (2015) 467–487.
[27] M. Hahsler, S. Chelluboina, K. Hornik, C. Buchta, The arules R-package ecosystem: analyzing interesting patterns from large transaction datasets, J. Mach. Learn. Res. 12 (2011) 1977–1981 URL: http://jmlr.csail.mit.edu/papers/v12/hahsler11a.html.
[28] O. Kucuktunc, B.B. Cambazoglu, I. Weber, H. Ferhatosmanoglu, A large-scale sentiment analysis for Yahoo! answers, Proceeding of the Fifth ACM International Conf. on Web Search and Data Mining (WSDM '12), New York, NY, USA, ACM, 2012, pp. 633–642.
[29] R. Jongeling, S. Datta, A. Serebrenik, Choosing your weapons: on sentiment analysis tools for software engineering research, Proceeding of Int'l Conf. on Software Maintenance and Evolution (ICSME'15), Bremen, IEEE, 2015, pp. 531–535.
[30] R. Jongeling, P. Sarkar, S. Datta, A. Serebrenik, On negative results when using sentiment analysis tools for software engineering research, Empirical Softw. Eng. J. 22 (5) (2017) 2543–2584.
[31] W. Maalej, Z. Kurtanovic, H. Nabil, C. Stanik, On the automatic classification of app reviews, Requirements Eng. 21 (3) (2016) 311–331.
[32] M.V. Mäntylä, N. Novielli, F. Lanubile, M. Claes, M. Kuutila, Bootstrapping a lexicon for emotional arousal in software engineering, Proceedings of the 14th

International Conference on Mining Software Repositories (MSR '17), Piscataway, NJ, USA, IEEE Press, 2017, pp. 198–202.

[33] S. Menard, Applied Logistic Regression Analysis, Applied Logistic Regression Analysis, Sage, Thousand Oaks, CA, 1995.

[34] Meta, —Reputation for comments? Last accessed: June http://meta.stackexchange.com/questions/296/reputation-for-comments, (2017) Last accessed: June.

[35] Meta – Stack exchange is too harsh to new users. Retrieved at http://meta.stackexchange.com/questions/179003/stack-exchange-is-too-harsh-to-new-users-please-help-them-improve-low-quality-po. Last accessed: June 2017.

[36] Meta—The NEW 'Be Nice' Policy. Retrieved at: http://meta.stackexchange.com/questions/240839/the-new-new-be-nice-policy-code-of-conduct-updated-with-your-feedback. Last accessed: June 2017.

[37] T. Mitra, E. Gilbert, The language that gets people to give: phrases that predict success on kick starter, Proc. of the 17th ACM Conf. on Computer Supported Cooperative Work & Social Computing (CSCW '14), ACM, New York, NY, USA, 2014, pp. 49–61.

[38] S.C. Müller, T. Fritz, Stuck and frustrated or in flow and happy: sensing developers' emotions and progress, Proc. of the 37th International Conf. on Software Engineering - Volume 1 (ICSE '15), IEEE, 2015, pp. 688–699.

[39] A. Murgia, P. Tourani, B. Adams, M. Ortu, Do developers feel emotions? An exploratory analysis of emotions in software artifacts, Proc. of the 11th Working Conf. on Mining Software Repositories, ACM, 2014, pp. 262–271.

[40] P. Nakov, A. Ritter, S. Rosenthal, V. Stoyanov, F. Sebastiani, SemEval-2016 task 4: sentiment analysis in Twitter, Proceedings of the 10th International Workshop on Semantic Evaluation, SemEval '16, June, San Diego, California, Association for Computational Linguistics, 2016.

[41] N. Novielli, F. Calefato, F. Lanubile, Towards discovering the role of emotions in stack overflow, Proc. 6th Int'l Workshop on Social Software Engineering (SSE'14), Nov. 16, 2014, Hong Kong, China, 2014, pp. 33–36.

[42] N. Novielli, F. Calefato, F. Lanubile, The challenges of sentiment detection in the social programmer ecosystem, Proc. of the 7th Int'l Workshop on Social Software Engineering (SSE '15), ACM, 2015, pp. 33–40.

[43] A. Ortony, G.L. Clore, A. Collins, The Cognitive Structure of Emotions, Cambridge University Press, 1990.

[44] M. Ortu, B. Adams, G. Destefanis, P. Tourani, M. Marchesi, R. Tonelli, Are bullies more productive? Empirical study of affectiveness vs. issue fixing time, Proc. of the 12th Working Conf. on Mining Software Repositories (MSR '15), Piscataway, NJ, USA, IEEE Press, 2015, pp. 303–313.

[45] J.W. Osborne, Bringing balance and technical accuracy to reporting odds ratios and the results of logistic regression analyses, Pract. Assess. Res. Eval. 11 (2006) 7.

[46] B. Pang, L. Lee, Opinion mining and sentiment analysis, Found. Trends Inf. Retrieval 2 (1–2) (2008) 1–135.

[47] S. Panichella, A. Di Sorbo, E. Guzman, C.A. Visaggio, G. Canfora, H.C. Gall, How can i improve my app? Classifying user reviews for software maintenance and evolution, Proceedings of the 2015 IEEE International Conference on Software Maintenance and Evolution (ICSME '15), 2015, pp. 281–290 DOI=http://dx.doi.org/10.1109/ICSM.2015.7332474.

[48] D. Pletea, B. Vasilescu, A. Serebrenik, Security and emotion: sentiment analysis of security discussions on GitHub, Proceedings of the 11th Working Conference on Mining Software Repositories (MSR 2014), New York, NY, USA, ACM, 2014, pp. 348–351.

[49] L. Ponzanelli, A. Mocci, A. Bacchelli, M. Lanza, Understanding and classifying the quality of technical forum questions, Proc. of the 14th International Conference on Quality Software (QSIC), 2014, 2014, pp. 343–352.

[50] M.M. Rahman, C.K. Roy, and I. Keivanloo. Recommending insightful comments for source code using crowdsourced knowledge. To appear in Proc. of the IEEE 15th Int'l Working Conf. on Source Code Analysis and Manipulation (SCAM 2015).

[51] M.R. Islam, M.F. Zibran, Leveraging automated sentiment analysis in software engineering, Proceedings of the 14th International Conference on Mining Software Repositories (MSR '17), Piscataway, NJ, USA, IEEE Press, 2017, pp. 203–214.

[52] Regression Diagnostics, In Research Methods II: Multivariate Analysis, Chapter 5, J. Tropical Pediatrics (2015) 44–55 Retrieved at: http://www.oxfordjournals.org/our_journals/tropej/online/ma.html , Last accessed: December 2015.

[53] J.A. Russell, A circumplex model of affect, J. Pers. Soc. Psychol. 39 (6) (1980) 1161.

[54] K.R. Scherer, T. Wranik, J. Sangsue, V. Tran, U. Scherer, Emotions in everyday life: probability of occurrence, risk factors, appraisal and reaction patterns, Soc. Sci. Inf. 43 (4) (2004) 499–570.

[55] SEmotion '16, Proceedings of the 1st International Workshop on Emotion Awareness in Software Engineering, New York, NY, USA, ACM, 2016.

[56] L. Singer, F. Figueira Filho, B. Cleary, C. Treude, M-A. Storey, K. Schneider, Mutual assessment in the social programmer ecosystem: an empirical investigation of developer profile aggregators, Proc. of the 2013 Conference on Computer supported cooperative work (CSCW '13), New York, NY, USA, ACM, 2013, pp. 103–116.

[57] J. Skeet, Writing the perfect question, http://tinyurl.com/stack-hints, (2010) (Last Accessed: June 2017).

[58] Stack Overflow - Be nice, Retrieved at: http://stackoverflow.com/help/be-nice (Last accessed: June 2017).

[59] Stack Overflow – How do i ask a good question? Retrieved at: http://stackoverflow.com/help/how-to-ask(Last accessed: March 2016).

[60] M-A. Storey, C. Treude, A. van Deursen, L-T. Cheng, The impact of social media on software engineering practices and tools, Proceedings of the FSE/SDP workshop on Future of software engineering research (FoSER '10), New York, NY, USA, ACM, 2010, pp. 359–364.

[61] M-A. Storey, The evolution of the social programmer, Proc. of the 9th IEEE Working Conf on Mining Software Repositories (MSR '12), Piscataway, NJ, USA, IEEE Press, 2012,140–140.

[62] M-A. Storey, L. Singer, B. Cleary, F. Figueira Filho, A. Zagalsky, The (R) evolution of social media in software engineering, Proc. of the on Future of Software Engineering (FOSE '14), New York, NY, USA, ACM, 2014, pp. 100–116.

[63] M. Thelwall, K. Buckley, G. Paltoglou, Sentiment strength detection for the social web, J. Am. Soc. Inf. Sci. Technol. 63 (1) (2012) 163–173.

[64] C. Treude, O. Barzilay, M-A. Storey, How do programmers ask and answer questions on the web? (NIER track), Proc. of the 33rd Int'l Conf. on Software Engineering (ICSE '11), New York, NY, USA, ACM, 2011, pp. 804–880.

[65] C. Wohlin, P. Runesson, M. Höst, M.C. Ohlsson, B. Regnell, A. Wesslén, Experimentation in Software Engineering. An Introduction, Kluwer Academic Publishers, 2000.

[66] D. Ye, Z. Xing, N. Kapre, The structure and dynamics of knowledge network in domain-specific Q&A sites: a case study of stack overflow, Empirical Softw. Eng. 22 ((February)1) (2017) 375–406.